\documentclass[aps,prb,twocolumn,superscriptaddress,a4paper,floatfix]{revtex4}

\usepackage[T1]{fontenc}

\usepackage[utf8]{inputenc}
\usepackage{amsfonts}
\usepackage{color}

\usepackage{dcolumn} 

\usepackage[pdftex]{graphicx}\pdfcompresslevel=9
\usepackage[pdftex,breaklinks=true,colorlinks=true]{hyperref}

\usepackage{ifthen}
\usepackage{ulem}

\newcommand{\journal}[4]
{\ifthenelse{\equal{#1}{prl}}{
\prl {\bf #2}, \href{http://link.aps.org/abstract/PRL/v#2/e#3}{#3} (#4)}
{\ifthenelse{\equal{#1}{prb}}{
\prb {\bf #2},
\href{http://link.aps.org/doi/10.1103/PhysRevB.#2.#3}{#3} (#4)}
{\ifthenelse{\equal{#1}{arxiv}}{preprint
\href{http://arxiv.org/abs/#2.#3}{arXiv:#2.#3}}
{\ifthenelse{\equal{#1}{rmp}}{
\rmp {\bf #2}, \href{http://link.aps.org/abstract/RMP/v#2/e#3}{#3} (#4)}
{\ifthenelse{\equal{#1}{cond-mat}}{preprint
\href{http://arxiv.org/abs/cond-mat/#2}{cond-mat/#2}}
{\ifthenelse{\equal{#1}{pre}}{
\pre {\bf #2}, \href{http://link.aps.org/abstract/PRE/v#2/e#3}{#3} (#4)}
{#1 {\bf #2}, #3 (#4)}}}}}}}

\newcommand{\journaldoi}[5]{#1\ {\bf #2}, \href{http://dx.doi.org/#5}{#3} (#4)}

\begin{document}

\title{Schwinger boson mean field theory: numerics for the energy landscape and gauge excitations
in two-dimensional antiferromagnets}
\author{G. Misguich}
\affiliation{
Institut de Physique Th\'eorique,
CEA, IPhT, CNRS, URA 2306, F-91191 Gif-sur-Yvette, France.}

\begin{abstract}
We perform some systematic numerical search for Schwinger boson mean field states  on square and triangular lattice clusters. We look for possible inhomogeneous ground states as well as low-energy excited saddle points.
The spectrum of the Hessian is also computed for each solution.
On the square lattice we find gapless $U(1)$ gauge modes in the non-magnetic phase.
In the $\mathbb{Z}_2$ liquid phase of the triangular lattice we identify the topological degeneracy as well as vison states.
\end{abstract}

\maketitle

\section{Introduction}

Quantum antifferomagnets 
can display complex many body phenomena, with rich phase diagrams, exotic states of matter with emerging degrees of freedom.\cite{balents10}
Indeed, minimizing a rather simple looking interaction like the Heisenberg one,
$\vec S_i\cdot\vec S_j$, can lead to a vast variety of states of matter, depending
on the size $\vec S_i^2=S(S+1)$ of the spins, the space dimension, geometry of the lattice, or relative strengths of possible competing interactions (frustration, etc.). In fact, after many years of theoretical investigations, the nature
of the ground state of the spin-$\frac{1}{2}$ Heisenberg model remains controversial on several two and three dimensional lattices. Some of the most interesting states that can be stabilized at
$T=0$ are called {\it spin liquids} and have no direct classical analogs. These systems remain rotationally invariant
down to zero temperature. There is a whole zoo of possible spin liquids, and
the most exotic one have low energy excitations which carry a half-odd-integer spin (unlike conventional spin waves).

There are rather few theoretical tools that are able to describe the limit of strong quantum fluctuations (small
$S$) for these systems, and spin liquid phases in particular. Among them, the so-called large-$N$ techniques, introduced by Affleck for spin chains,\cite{a85} play a central role. By generalizing the symmetry group of global spin rotations
from SU(2) to some larger group like SU($N$) or Sp($2N$),\cite{rs91} the model can  be solved in the limit $N=\infty$.
Of course, the physics for $N=2$ needs not be simply related to that at $N=\infty$. However, the
success of these approaches is in part due to the fact that a number of interesting states that can be realized
for SU(2) models, also exist in the $N=\infty$ phase diagrams. 
In particular, the $N=\infty$ models are not restricted to have magnetically ordered ground state, and  can have resonating valence-bond spin liquids ground states
with fractionalized excitations and emerging gauge degrees of freedom.
Furthermore, the effect of finite-$N$ corrections (both perturbative\cite{r90,tmgc97} and non-perturbative ones\cite{rs89}) can be addressed.

In this work we consider a particular large-$N$ limit, the so-called Schwinger boson mean field theory (SBMFT).\cite{aa88,rs91}
It has been applied to
many two-dimensional systems, such as  the $J_1-J_2$ square,\cite{mpb91} triangular,\cite{sachdev92,gc93,cmm01,msmt11} honeycomb\cite{mfe94,wang2010},
kagome,\cite{sachdev92,wv06,tm11,mbl12} kagome with further neighbor interactions,\cite{mbl12,fak12}  or Shastry-Sutherland\cite{am96,cms01} or CaV$_4$O$_9$,\cite{am96b} lattices.
This limit can in particular stabilize magnetically ordered (N\'eel) states as well as $\mathbb Z_2$ spin liquids
with gapped spinons.\cite{sachdev92}

The model contains $N$ flavors of spin-$\frac{1}{2}$ bosons (spinons) and the parameter which plays the role of the spin value $2S$ is the number $\kappa$ of bosons per site (and per flavor).
After performing a Hubbard-Stratonovich decoupling of the boson-boson interaction, the different boson flavors are coupled to a single complex field $A_{ij}$ on each bond $ij$ of the lattice.
The formal (Gaussian) integration of the spinons gives an effective action $S_{{\rm eff}, N}[A]$ for the bond field.
But the dependence in $N$ is simple since this action is that of the single-flavor problem multiplied by a factor $N$:
$S_{{\rm eff}, N}[A]=NS_{\rm eff, 1}[A]$.
When $N$ is large it is therefore natural to perform a saddle point expansion.
At the saddle point the bond-field fluctuations are frozen, and the effective Hamiltonian  is simply quadratic in the spinon operators.
So the action can be computed with the help of a standard Bogoliubov transformation.
Finding a the large-$N$ ground state thus amounts to solving a classical minimization problem.

Previous SBMFT studies have mostly been  focused on the ground state properties at $N=\infty$.
However,  as a first step to understand the effects of finite-$N$ corrections, it is interesting to have access
to the low-energy energy landscape of saddle points. Although the system is locked into the lowest one when $N=\infty$, the (large but) finite-$N$
physics must include some fluctuations between different  low-energy  saddle points, as well as some (perturbative in $1/N$) fluctuations in the vicinity of each of these saddle points.
Our goal here is to provide a quantitative description of some excited saddle points. 

In order to reduce the number of variables, it is almost always assumed that the lowest-energy
state preserves most (or at least some) of the lattice symmetries (see Ref.~\cite{hgv09} for a notable exception).
In practice most numerical studied have so far been restricted to solutions with a unit cell including a few sites only.
This is a reasonable assumption for the ground state, and we confirm validity in several cases.
However, since we are interested here in excited saddle points, we need to be able to compute some {\it spatially inhomogeneous} states as well. We will however limit ourselves to time-independent solutions.

In this work perform some extensive numerical minimization to look for
the ground state and low-energy excited states on clusters containing up to 144 sites. We compare the results we obtain on two different lattices (square and  triangular) and
for different values of the boson density (``spin'') $\kappa$.

\section{Schwinger boson mean field theory}
\label{sec:sbmft}

To keep this article self-contained, this section presents the basic ideas and notations of the SBMFT (see also Refs.~\cite{auerbachbook,misguich08}).

``up'' and ``down'' bosons operators ($\sigma\in\{\uparrow,\downarrow\}$), carrying a $S=1/2$, are introduced at
each lattice site: $b^\dag_{i\sigma}$ and $b_{i\sigma}$. 
The spin operators can then be written: 
\begin{eqnarray}
S^+_i&=&b^\dag_{i\uparrow}b_{i\downarrow}\\
S^-_i&=&b^\dag_{i\downarrow}b_{i\uparrow}\\
2S^z_i&=&b^\dag_{i\uparrow}b_{i\uparrow}-b^\dag_{i\downarrow}b_{i\downarrow}
\end{eqnarray}
These relations imply that  the commutation relations
$[S^\alpha_i,S^\beta_i]=i\epsilon^{\alpha\beta\delta}S^\delta_i$
are automatically verified. The total spin reads
$
\vec S_i^2=\frac{n_i}{2}\left(\frac{n_i}{2}+1\right)
$,
where $n_i=b^\dagger_{i\uparrow}b_{i\uparrow}
+b^\dagger_{i\downarrow}b_{i\downarrow}$ is the total number of bosons at site $i$.
To fix the ``length'' of the spins, the following constraint must therefore be imposed on physical states: 
\begin{equation}
\label{eq:constraint}
\forall i,\quad \sum_\sigma b^\dag_{i\sigma}b_{i\sigma}=\kappa=2S 
\end{equation}

The Heisenberg exchange Hamiltonian is biquadratic in the $b$ operators and reads
\begin{eqnarray}
H&=&\sum_{\langle ij\rangle}J_{ij}\,\mathbf S_i \cdot\mathbf S_j\\
&=&\frac{1}{4}\sum_{\langle ij\rangle}J_{ij}\, \left( b_{i\sigma}^\dag \vec{\sigma}_{\sigma,\sigma'} b_{i\sigma'}\right)\cdot\left(b^\dag_{j\tau}\vec\sigma_{\tau,\tau'}b_{j\tau'}\right), 
\end{eqnarray}
where $\vec \sigma$ is the vector whose components are the Pauli matrices, and each lattice bond $ij$ is taken only once. 
It is convenient to re write $H$ using rotationally invariant bond operators:
\begin{equation}
\label{eq:HAB}
H=\sum_{\langle ij\rangle}J_{ij}\,  \left(:\hat B_{ij}^\dag\hat B_{ij}:-\hat A_{ij}^\dag\hat A_{ij}\right),
\end{equation}
where $:-:$ represents normal ordering and the  bond operators $\hat A_{ij}$ and $\hat B_{ij}$ are 
\begin{eqnarray}
\hat A_{ij}&=&\frac{1}{2}(b_{i\uparrow} b_{j\downarrow}-b_{i\downarrow} b_{j\uparrow})\label{eq:defA}\\
\hat B_{ij}&=&\frac{1}{2}(b_{i\uparrow}^\dag b_{j\uparrow}+b_{i\downarrow}^\dag b_{j\downarrow}). \label{eq:defB}
\end{eqnarray}
$\hat A^\dag_{ij}$ creates a spin singlet on the (oriented) bond $ij$ whereas $\hat B^\dag_{ij}$ creates a triplet.
Due to the constraints, these operators are linked by the relation 
\begin{equation}
 :\hat B_{ij}^\dag\hat B_{ij}:+\hat A_{ij}^\dag\hat A_{ij}=S^2 
\end{equation}
and it is therefore possible to express the Hamiltonian using $\hat A$ only:
\begin{equation}
H= \sum_{\langle ij\rangle}J_{ij}\,  \left(S^2-2\hat A_{ij}^\dag\hat A_{ij}\right).
\end{equation}

The SB mean field approximation -- which can be formally justified in a large-$N$ limit of the model-- consists in decoupling the quartic terms and to
add some chemical potentials $\lambda_i$ to tune
the {\it average} number of boson at each site site (instead of Eq.~\ref{eq:constraint}).
The resulting mean field Hamiltonian is
\begin{eqnarray}
\label{eq:HA}
H_{\rm MF}&=&\sum_{\langle ij\rangle}J_{ij} \left(S^2
-2 \overline{A_{ij}}\hat A_{ij}-2  A_{ij}  \hat A_{ij}^\dag +2 |A_{ij}^2|
\right)  \nonumber\\
&&-\sum_i \lambda_i\left(b_{i\sigma}^\dag b_{i\sigma}-\kappa
\right)
\end{eqnarray}
where $\overline x$ is the conjugate of $x$, $A_{ij}$ are complex link variables with property $A_{ij}=-A_{ji}$.
It is also possible to write a mean field theory keep simultaneously both operators $\hat A$ and $\hat B$ on each bond.\cite{cgt93,fc09} This will however not be considered here.
\cite{noteSpn}

It is convenient to write $H_{\rm MF}$ by grouping the creation and annihilation operators in a vector
$\hat\phi=(b_{1\uparrow},\cdots,b_{N_S\uparrow},b_{1\downarrow}^\dagger,\cdots, b_{N_S\downarrow}^\dagger)^t$, where $N_S$ is the number of sites in the lattice. Eq.~\ref{eq:HA} then reads:
\begin{equation}
 H_{\rm MF}=(\hat \phi)^\dagger M \hat\phi + {\rm cst.}
\label{eq:HM}
\end{equation}
where the $M$ is a $2N_S\times2N_S$ matrix :
\begin{equation}
 M=\left[
\begin{array}{cc}
-\lambda_i & J_{ij} A_{ij} \\
-J_ {ij}\bar A_{ij}&  -\lambda_i
\end{array}
\right].\label{eq:M}
\end{equation}
For simplicity, we restrict the discussion to finite systems where $H_{\rm MF}$ has a gapped spectrum (possibly vanishing in the thermodynamic limit).
The ground state exists and its spectrum is gapped if and only if the the matrix $M$ is positive definite.\cite{colpa78}
We define a diagonal $2N_S\times2N_S$ matrix $\sigma$:
$ \sigma=\left[
\begin{array}{cc}
-1 & 0 \\
 0 &  1 \end{array}
\right]$
and, with the gap condition, $\sigma M$ is diagonalizable (although not Hermitian) and has pairs of real eigenvalues $\pm \omega_{n=1\cdots N_S}$, where
the $\omega_n>0$ are the of the Bogoliubov modes of $H_{\rm MF}$.
The ground state energy of $H_{\rm MF}$ is given by
\begin{equation}
 E_{\rm MF}=\sum_n \omega_n+\sum_{ij}J_{ij}|A_{ij}|^2+(2S+1)\sum_i \lambda_i 
\end{equation}
(remark: in a large-$N$ formalism, this corresponds to the energy {\it per flavor}).

The self consistency is reached when $A_{ij}=<\hat A_{ij}>_{GS}$, which is equivalent to
$\partial E /\partial A_{ij}=0$, where $E$ is the energy of the  ground state of $H_{\rm MF}$.
The average number of boson
per site is equal to $2S$ (Eq.~\ref{eq:constraint}) at the point(s) where $\partial E /\partial \lambda_i=0$.
Since $\partial^2 E /\partial \lambda^2\leq 0$ (see this footnote\cite{note2}),
 this point corresponds to
a {\it maximum} of $\left\{\lambda_i\right\}\mapsto E[\left\{A_{ij}\right\},\left\{\lambda_i\right\}]$.
For fixed $\left\{A_{ij}\right\}$, this allows to use a maximization algorithm to determine the chemical potentials $\lambda_i[\left\{A_{ij}\right\}]$.
In a similar way, the self-consistent $\left\{A_{ij}\right\}$ correspond to a {\it minimum} of $E\left[\left\{A_{ij}\right\},\lambda_i[\left\{A_{ij}\right\}]\right]$.

\subsection{Gauge invariance and fluxes}

The solution of this system of equations is not unique, at least because of the $U(1)$ gauge invariance of the Hamiltonian. 
Under the gauge transformation 
\begin{eqnarray}
b_{j\sigma}&\to& e^{i\theta(j)}b_{j\sigma} \nonumber\\
\hat A_{ij}&\to& e^{i(\theta(i)+\theta(j))} \hat A_{ij}
\label{eq:GT}
\end{eqnarray}
the physical operators (as $\vec S_i$ or $H$) stay unchanged.
In other words, the bond parameters $A_{ij}$ label the physical state in a redundant way.
Using gauge transformations we can fix the phases of some $A_{ij}$, without changing any physical observable.
This decreases the number of variables to be optimized, and is therefore useful in numerical studies.

The moduli $|A_{ij}|$ are of course gauge invariant quantities,  simply related to the energy.
But some combinations of complex phases around closed loops are also gauge invariant.
Consider the following  operator $\hat \mathbb A_{i_1 i_2 \cdots i_{2n}}$:
\begin{equation}
\hat \mathbb A_{i_1 i_2 \cdots i_{2n}}=\hat A_{i_1 i_2}(-\hat A_{i_2 i_3 }^\dag)\hat A_{i_3 i_4}...(-\hat A_{i_{2n} i_{1}}^\dag).
\end{equation}

This operator is defined on any loop with an even length on the lattice and is manifestly gauge invariant. Its  mean field (or large-$N$) counter part
\begin{equation}
\mathbb A_{i_1 i_2 \cdots i_{2n}}=A_{i_1 i_2}(-\bar A_{i_2 i_3 })A_{i_3 i_4}...(-\bar A_{i_{2n} i_{1}})
\end{equation}
gives the flux $\phi_{i_1 i_2 \cdots i_{2n}}=\arg (\mathbb A_{i_1 i_2 \cdots i_{2n}})$
recently discussed by Tchernyshyov {\it et al.}\cite{tms06}

In general, the number of variables (all the moduli, plus some phases) grows  with the system size. Since
it is quite difficult to perform an exhaustive search for mean field solutions if the number of parameters
is extensive, the usual strategy is to decrease the number of bond variables by {\it assuming} that the ground state solution preserves some (or all) symmetries of the lattice.
There are  however some examples where the lowest energy solution spontaneously {\it breaks} some symmetries (kagome clusters with 36 or 48 sites\cite{gmSBMFT}).

\section{Numerical method}

We describe an algorithm to find saddle points (self-consistent mean field state),
local minima, and (hopefully) global energy minima.

\subsection{Determination of the chemical potentials}\label{ssec:dcp}
For given values of the bond parameters $A_{ij}$, one should first 
adjust the chemical potentials ($\lambda_i$).

The basic idea is to perform a (non-linear) least-square minimization of
$\sum_i f_i^2$, where $f_i(\{\lambda_i\})=\langle\hat n_i\rangle-\kappa$
is a function of the chemical potentials.
Since each density $\langle\hat n_i\rangle$ is an increasing function
of $\lambda_i$, this method converges relatively rapidly.
To do so, we use an implementation of the Levenberg-Marquart algorithm.\cite{levmar}
To make the method faster, we provide the gradient matrix
$G_{ij}=\frac{\partial \langle \hat n_i \rangle}{\partial\lambda_j}$ explicitly.
$G_{ij}$ is easy to compute using linear-response theory (using the matrix which implements the Bogolibov transformation,
as well as the energies $\omega_i$).
The iterative minimization should start in a region of the space of $\lambda$ where
the Bogoliubov transformation exist ($M>0$). If the  values of $\lambda$ obtained at the previous step
do not satisfy this condition with the new $\{A_{ij}\}$, one starts the least-square minimization of $\sum_i f_i^2$
from a sufficiently low and uniform $\lambda$.

Finally we note that for some values of the bond parameters $A_{ij}$, there
is no uncondensed state satisfying $\langle\hat n_i\rangle=\kappa$.
If such a situation is encountered, we add some artificial energy penalty so that the energy-minimization algorithm (next section) tends to escape this point.

\subsection{Optimization of the bond parameters $A_{ij}$}\label{ssec:obp}

Equipped with a procedure to compute the $\lambda_i$ as a function of the $A_{ij}$,
we can start to look values of $A_{ij}$ which correspond to self-consistent mean field states.
The first stage is simply to iterate the bond self-consistency conditions in
the usual way:
\begin{itemize}
 \item[0] We start from initial random bond values (or perturbing a previously found solution).
  \item[1] After adjusting the $\lambda$ by the procedure above, the  ground state
is obtained by Bogoliubov transformation, and the  expectation values $\langle A_{ij} \rangle$ are computed.
\item[2] The bond parameters are replaced by the values above. 
The new  bond parameters are gauge-transformed to a fixed gauge choice where a maximum number of bond parameters are set to be real.
This avoids some possible slow drift of some complex phases of the $A_{ij}$
which would be non-physical.
\item[3] Go back to step 2 until the bond parameters do not change by more than a small
threshold $\epsilon$. 
\end{itemize}
This method was used by Hermele {\it et al.}\cite{hgv09} in a similar context.
It is easy to check that local minima are attractive for these iterations while local maxima a repulsive.
More precisely, the error will decrease (resp. increase) in the directions corresponding to
eigenvectors of the Hessian (anticipating on Eq.~\ref{eq:Hess}) with positive (resp. negative) values. The convergence is faster if the Hessian eigenvalues are large and positive.
It practice, the iterations above allows to quickly go down in energy, but it is
not  efficient to achieve a full convergence when the system has more than $\sim$ 10 bonds or so.
Indeed, it turns out that a high accuracy is required to resolve the possibly small  differences (bond modulations, etc) between different saddle points.

To find accurately self-consistent state, the second stage amounts to perform a least-square
optimization on $\sum_{ij} g_{ij}^2$, where $g_{ij}(\{A\})=\langle\hat A_{ij}\rangle-A_{ij}$
is a function of the bond parameters. Of course, the expectation values $\langle\hat A_{ij}\rangle$
are computed after the  $\lambda_i(\{A\})$ have been determined.
This second stage allows to converge not only to local extrema but to saddle points as well.
Again, we use a Levenberg-Marquart minimization algorithm,\cite{levmar} with explicit calculation
of the gradient $\frac{\partial \langle\hat A_{ij}\rangle}{\partial A_{kl}}$ at fixed densities $\langle\hat n\rangle=\kappa$.
The later is again obtained using linear-response theory, with the additional
complexity that it contains some terms coming from the variation
of the $\lambda$ :
\begin{eqnarray}
 \left.
	\frac{\partial \langle\hat A_{ij}\rangle}{\partial A_{kl}}
\right|_{\langle\hat n\rangle=\kappa}
	&=& \left.
	\frac{\partial \langle\hat A_{ij}\rangle}{\partial A_{kl}}
\right|_{\lambda{\rm \;fixed}} \nonumber \\
&&
+ \sum_{a,b}\frac{\partial \langle\hat A_{ij}\rangle}{\partial \lambda_{a}}
\left[
	\frac{\partial \langle\hat n\rangle}{\partial \lambda}
\right]^{-1}_{ab}
\left.\frac{\partial \langle\hat n_b\rangle}{\partial A_{kl}}\right|_{\lambda{\rm \;fixed}}
\label{eq:dAdA}
\end{eqnarray}


As for the first stage, we work with bond variables corresponding to a fixed gauge choice. In particular, the matrix of Eq.~\ref{eq:dAdA} is evaluated in the subspace
of bond parameter variations which is orthogonal to pure gauge transformations.
This is important to get rid of unphysical slow phase drifts during the iterations. Most of the time a double precision accuracy in reached  in less than 10 Levenberg-Marquart iterations when the system has less than one hundred bonds or so.

Finally the optimization is repeated using at least  a few hundred (often thousands) of random initial conditions for the $A_{ij}$.
This is relatively time consuming since each evaluation of the energy (or its derivatives) requires a numerical adjustment of the chemical potentials (Sec.~\ref{ssec:dcp}), which is
itself a (convex) least-square problem with many variables.

Varying the parameter $\epsilon$ (stopping criterion for the iterations of stage 1) allows to tune if the method will converge toward very low-energy saddle points (small $\epsilon$),
or saddle point at higher energy (larger $\epsilon$).
If $\epsilon$ is very small (say $10^{-5}$) the first stage will terminate close to the ground state and the second step (least-square) will converge to the ground state with high probability if  the system is not too large ($\lesssim 100$ bonds).
On the other hand, if $\epsilon$ is too large, the first stage will stop at some relatively high energy configuration.
In such region of the $A$ space we expect a very (exponentially) high density of saddle points.
In practice this density of saddle point is so high that the program will find a new solution at every run, and a given saddle point will rarely be obtained twice.
The best choice is to adjust $\epsilon$ so that
the iteration stage leads to configurations in a typical energy range above the ground state where the number of saddle points is not too large (a few tens).
In such a case, after a sufficiently large number of runs, one obtains the full (or almost full) list of saddle points in that energy window. Of course, due to the large number of variables,
one cannot exclude the presence of some additional saddle points with a small basin of attraction with respect to this algorithm.

\subsection{Hessian and stability}

The least-square procedure described above leads to self-consistent mean field states, satisfying $A_{ij}=\langle \hat A_{ij} \rangle$.
These states are saddle point of the energy (considered as a function of the $A_{ij}$, and the chemical potential being themselves functions of the $A_{ij}$).
To check if each mean field state is a  local minimum, local maximum, or generic saddle points with stable as well as unstable directions, we compute the Hessian matrix:
\begin{equation}
K_{l,l'}=\frac{\partial ^2 E_{\rm MF}}{\partial A_l^\epsilon \partial A_{l'}^{\epsilon'}}
 =4J_l \left( \delta_{l l'}\delta_{\epsilon \epsilon'} - \left.\frac{\partial \langle \hat A_l^\epsilon \rangle}{\partial A_{l'}^{\epsilon'} }\right|_{\langle\hat n\rangle=\kappa} \right)
 \label{eq:Hess}
\end{equation}
where $l$ and $l'$ represent two bonds and $\epsilon$ and $\epsilon'$ denote the real or Imaginary part of the bond variables
(see Eq.~\ref{eq:dAdA} for derivatives of the bond expectation values at fixed densities).

Due to the gauge invariance of the model, the Hessian  contains some zero eigenvalues associated to infinitesimal gauge transformations. The number of such gauge modes
can be computed on each given lattice (using the rank of a modified adjacency matrix of the lattice\cite{messio10}),
and these non-physical zero eigenvalues are of course be discarded when discussing these solutions.
For the clusters studied here, the number of zero eigenvalues is always equal to the number of pure gauge modes, so we can conclude that there is no {\it physical} zero mode in the Hessian.
The sign of the smallest non-zero eigenvalue of $K$ tells us whether the mean field state is a local minimum, or an unstable saddle point.

The spectrum of the Hessian for the ground state gives some information about the magnitude of the $1/N$ corrections due to Gaussian fluctuations in the vicinity of the energy minimum.
Finally, the spatial structure of the lowest eigenvector of the Hessian gives some information about the physical nature of these fluctuations.

\section{Numerical results}

\subsection{Square lattice}

The SBMFT  phase diagram is well known on the square lattice:  magnetic long-range order for $\kappa\gtrsim 0.39$\cite{aa88}
and a disordered phase with gapped spinons for $\kappa\lesssim 0.39$ (``Coulomb'' phase, unstable at finite-$N$\cite{rs89}).
In the ordered phase the spinon gap drops as $\sim 1/N_s$  ($N_s$ is the number of sites) and Bose-condensation occurs in the thermodynamic limit, leading to spontaneous break down of the spin rotation symmetry. By considering here only finite clusters, the ground state is always rotationally invariant and the gap finite. Still, both phases can be distinguished using standard finite size-scaling for the gap or spin-spin correlations. 
We will discuss how the ``energy landscape'' of mean field saddle points differ between the two  phases.

\subsubsection{Hessian of the ground state and gauge modes}

We did some extensive search for saddle points on square lattices with  36 sites and with $\kappa=0.1$ and $\kappa=1$.
The ground state, as expected, is spatially uniform, real, and has a vanishing flux on all the square plaquettes, whatever the boson density.
However, the  spectrum of the Hessian is quite different in the magnetic phase and
in the disordered phase.

\begin{figure}
\includegraphics[angle=-90,width=6cm]{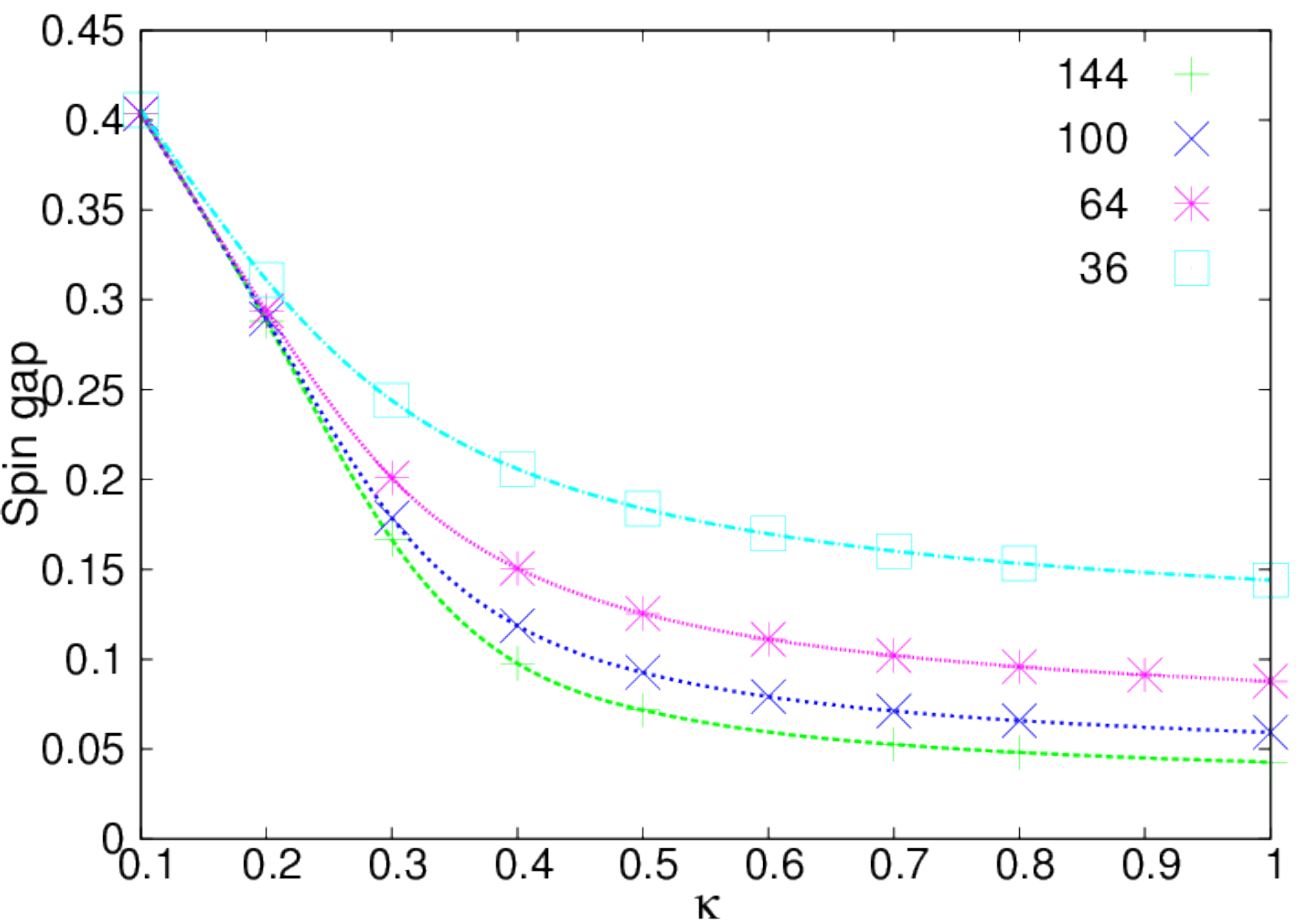}
\includegraphics[angle=-90,width=6cm]{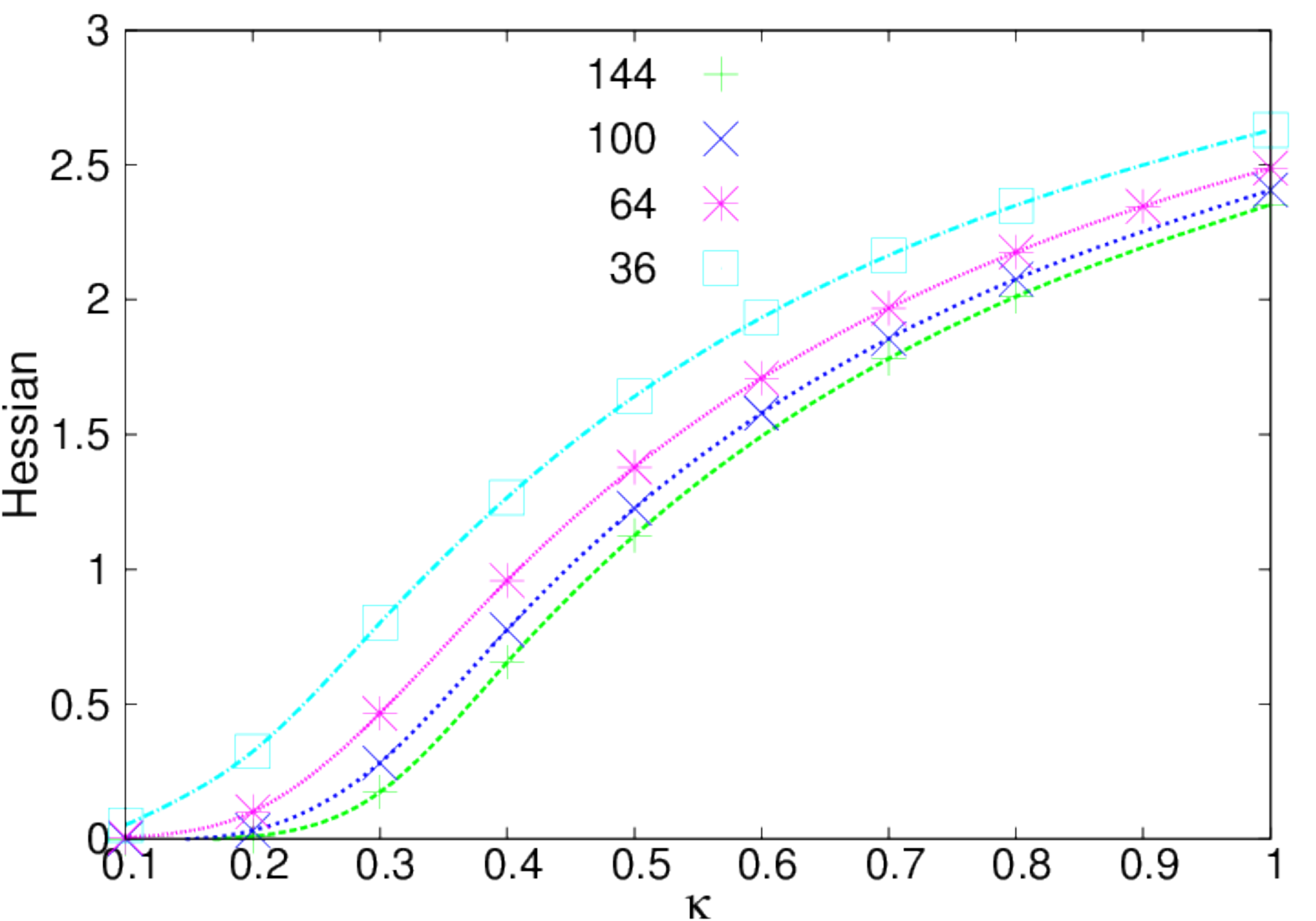}
\includegraphics[angle=-90,width=6cm]{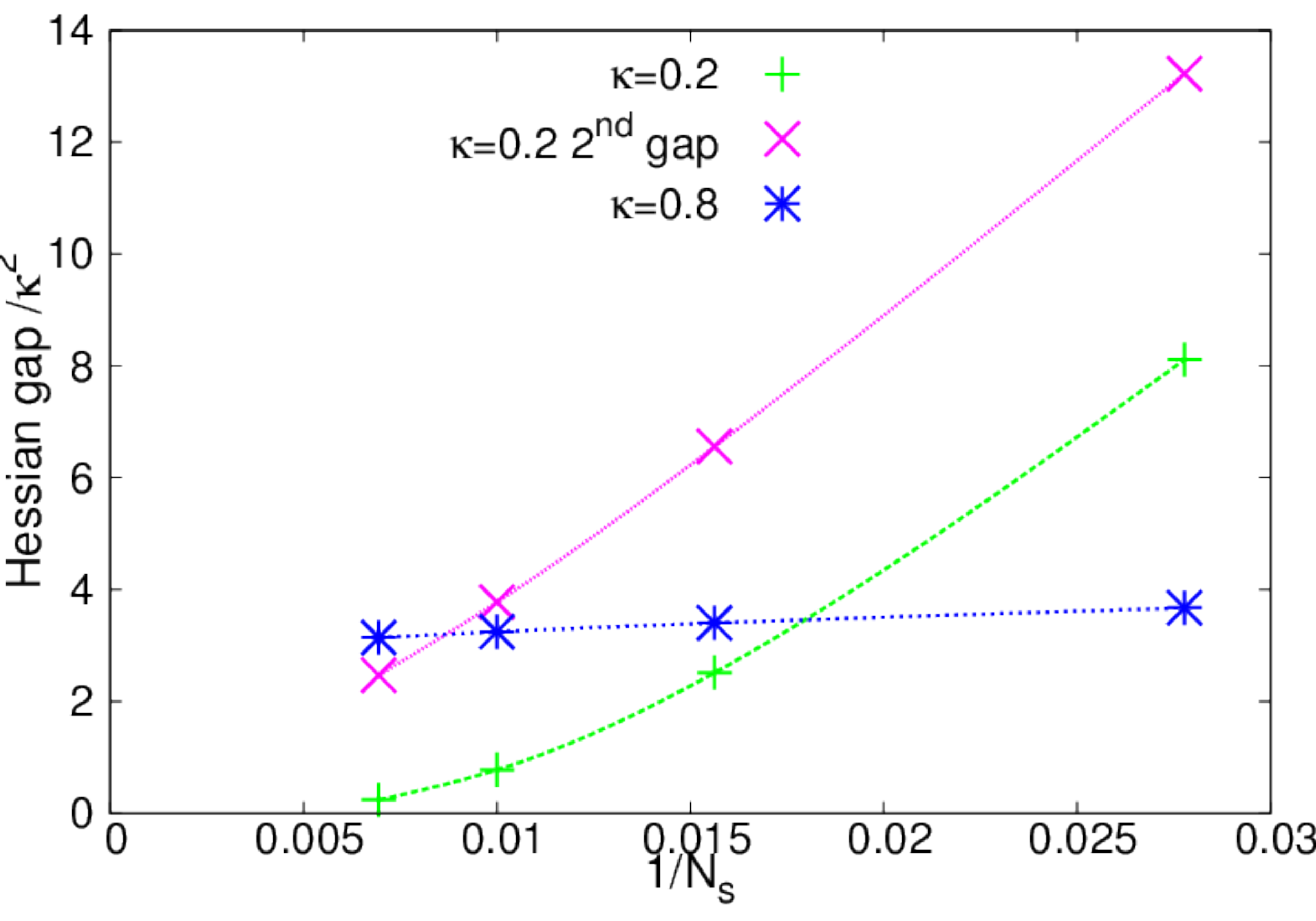}
\caption{(Color online) Square lattice model. Top: spin gap as a function of $\kappa$.
In the thermodynamic limit this state is associated to magnetic long-range order (vanishing spin gap) for $\kappa> \kappa_c\simeq 0.39$.\cite{aa88}
Middle panel: smallest eigenvalue of the Hessian for the ground state state on 36, 64 and 144-site square lattices.
Bottom: scaling of the lowest Hessian eigenvalue as a function of the system size $N_s$. This shows that the Hessian is gapless for $\kappa=0.2$ and gapped
for $\kappa=0.8$. The Hessian gap is finite in the magnetic phase, while it vanishes in the thermodynamic limit for $\kappa<\kappa_c$.
For $\kappa=0.2$, first eigenvalues decays faster than $1/N_s$  whereas the second eigenvalue  goes to zero as $\sim 1/N_s$ (see text).
}
\label{fig:gap_hess_sq}
\end{figure}

In Fig.~\ref{fig:gap_hess_sq} the smallest eigenvalue of the Hessian (Eq.~\ref{eq:Hess}) is plotted as a function of the $\kappa$ for different system sizes (up to 144 sites).
Although  finite-size effects are important, these data indicate  that the Hessian is gapped in the thermodynamic limit for large $\kappa$, while it becomes gapless for small $\kappa$.
The transition very likely coincides with that of magnetic long-range order.

The gaplessness of the Hessian in the disordered region is due to the bipartite character of the lattice.
On a bipartite lattice, the bond parameters are invariant under {\it staggered} gauge transformations: 
\begin{eqnarray}
b_{r\sigma}&\longrightarrow& b_{r\sigma} e^{i(-1)^r\theta } \\
A_{rr'}&\longrightarrow& A_{rr'}
\end{eqnarray}
which, in Wen's terminology,\cite{wen02} means that the invariant gauge group (IGG) is $U(1)$.

If  we perform a spatially varying gauge transformation including the staggered factor we get
\begin{eqnarray}
b_{r\sigma}&\longrightarrow& b_{r\sigma} e^{i(-1)^r\theta(r) } \\
A_{rr'}&\longrightarrow& A_{rr'}e^{i(\theta(r) - \theta(r'))}\;,
\label{eq:stagGT}
\end{eqnarray}
where $r$ (resp. $r'$) is on the even (resp. odd) sublattice. Starting from a state described by $A^0_{rr'}$
we construct a phase fluctuation of the bond parameters parametrized by $a_{rr'}\in\mathbb R$:
\begin{equation}
A_{rr'}'=A_{rr'}^0+dA_{rr'}=A_{rr'}^0e^{i a_{rr'}} \label{eq:arr}
\end{equation}
From Eq.~\ref{eq:stagGT},  $a_{rr'}$ transforms as a conventional $U(1)$ gauge field:
\begin{equation}
a_{rr'}\longrightarrow a_{rr'}+\theta(r) - \theta(r').\label{eq:agt}
\end{equation}
So, if we denote by $E(\{a_{rr'}\})$ the energy of the perturbed mean field state (after the appropriate adjustment of the chemical potentials),
the energy should be gauge invariant under Eq.~\ref{eq:agt}.

In the small-$\kappa$ phase where the bosons are gapped and their correlation length is short, $E(\{a_{rr'}\})$ should
be a local (short-ranged) function of the fluctuation $a_{rr'}$.

\begin{figure}
\includegraphics[width=8.3cm]{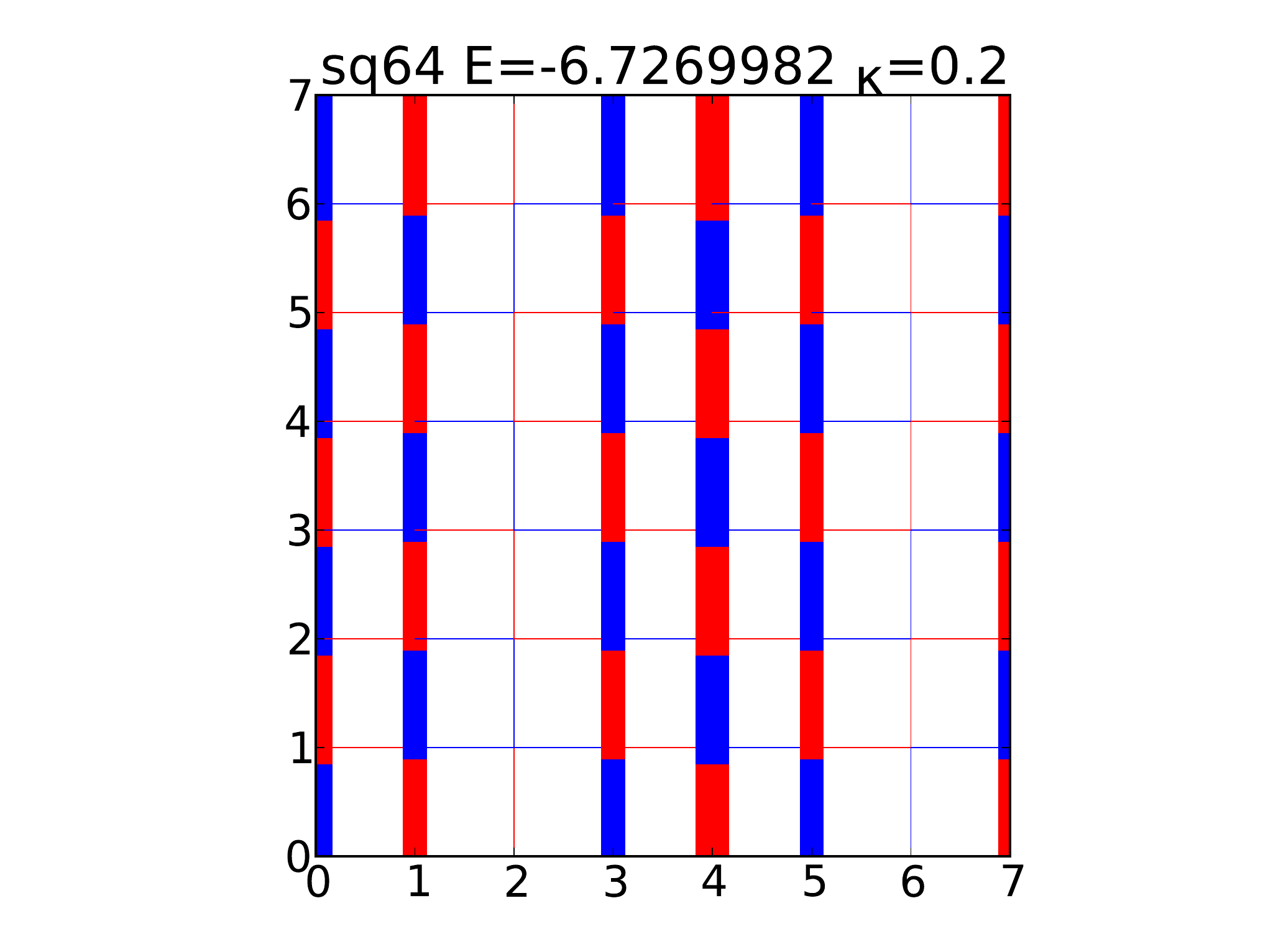}
\caption{(Color online) Hessian eigenvector $dA_l$ corresponding to the second smallest eigenvalue (0.26224396339) for a 64-site square lattice with $\kappa=0.2$.
The thickness of each bond $l$ is proportional to the modulus $|dA_l|$ while the color represents the complex argument of $dA_l/A_l$.
Red: argument is $+\pi/2$, blue is $-\pi/2$. Here $dA$ is a gauge excitation,
it corresponds to a long wavelength modulation of the flux through each plaquette. Here the wave-vector is parallel to the horizontal bonds.
This gauge mode is associated to the smallest non-zero wave vector on the square lattice and is therefore
four-fold degenerate.}
\label{fig:SecHessianModeSq64}
\end{figure}

The simplest local term compatible with gauge invariance would be the lattice version of the magnetic energy $( \vec\nabla \times \vec a)^2$,
and would take the form of the square of the circulation of $a$ around small loops
(this can also be obtained from a small-$\kappa$ expansion\cite{tms06}).
This would give an Hessian eigenvalue scaling as the square of the smallest available wave-vector, that is $\sim 1/L^2\sim 1/N_s$ ($L$ the linear size and $N_s$ the number of sites).
Such modes are indeed found in the spectrum of the Hessian and correspond to the {\it second} non-zero eigenvalue. The associated eigenvector is displayed in Fig.~\ref{fig:SecHessianModeSq64}.
The thickness of each bond $l$ is related to the modulus $|dA_l|$ while the color represents the complex argument of $dA_l/A_l$.
In the present case, these complex arguments take only two values: $\pm\pi/2$ (blue and yellow), indicating that $dA$ is a {\it gauge mode}
of the form of Eq.~\ref{eq:arr}, with $dA_l/A^0_l\sim i a_l\in i\mathbb{R}$. 
In the present case $a_l\sim \cos(2\pi x_l/L)$ for a vertical link $l$ (oriented from the even to the odd sublattice) at horizontal position $x_l$, and $a_l=0$ on horizontal bonds.
This gauge model may be viewed as a low-energy ``photon'' of the effective gauge theory.\cite{rs89}

\begin{figure}
\includegraphics[width=8.3cm]{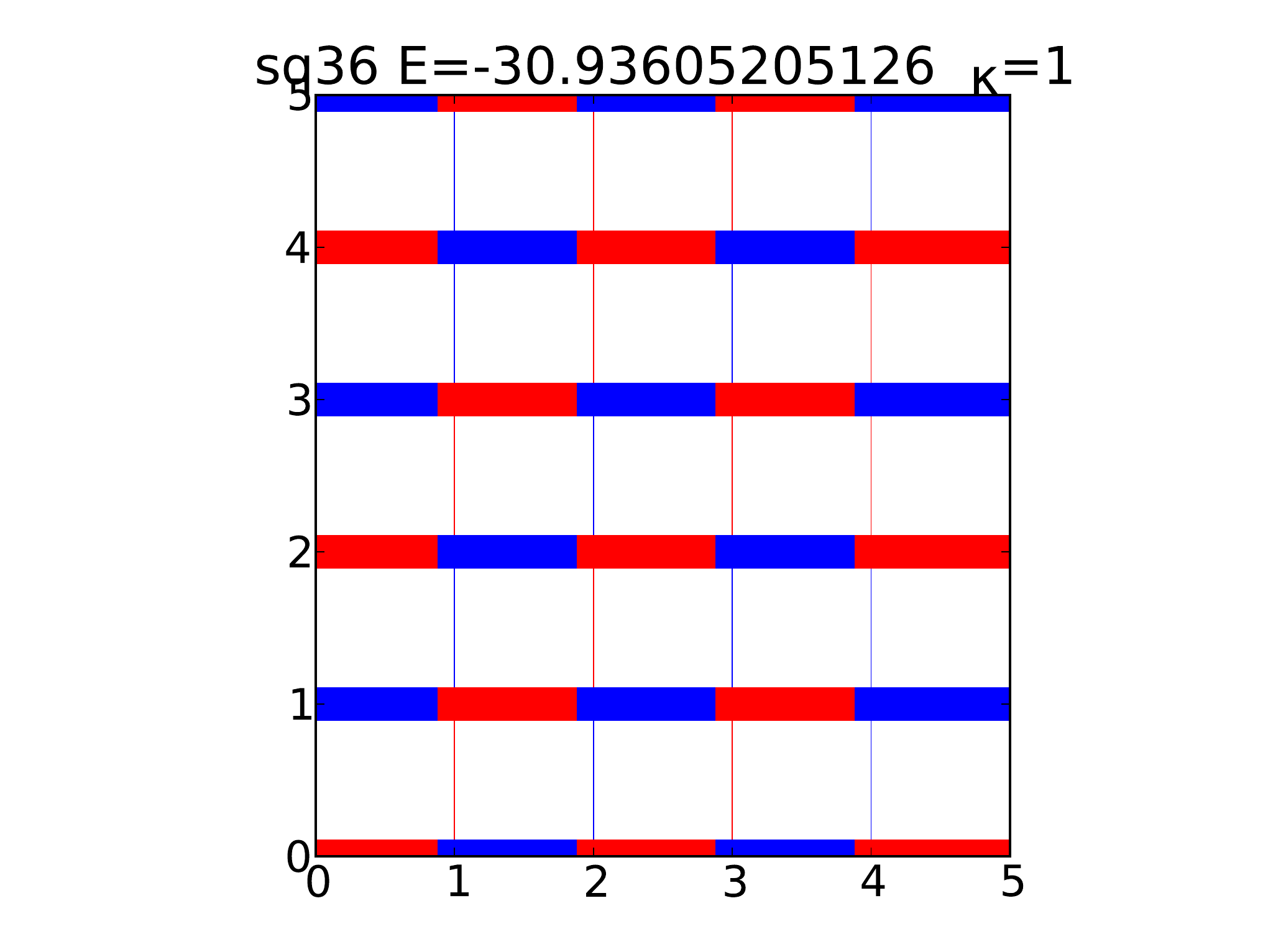}
\caption{(Color online) Hessian eigenvector $dA_l$ corresponding to the smallest eigenvalue (2.62955994344) for a 36-site square lattice.
Same representation as in Fig.~\ref{fig:SecHessianModeSq64}.
The complex argument  of $dA_l/A_l$ takes only two values: $\pm\pi/2$ (blue and red), indicating that $dA$ is a gauge excitation.
This mode corresponds to an increase of the flux through large horizontal loops, and no change for the flux going through local plaquettes.}
\label{fig:HessianModeSq36}
\end{figure}

In fact the numerical data indicate that the {\it lowest} eigenvalue of the Hessian decays {\it faster} than
$1/N_s$ (Fig.~\ref{fig:gap_hess_sq}). 
The associated mode is represented
in Fig.~\ref{fig:HessianModeSq36}, and is also a gauge mode ($dA_l/A_l^0$ is purely imaginary). Inspecting the sign of $a_l$
one sees that it corresponds to a change in the ``global'' flux associated to the large loops encircling the torus while local loops are unaffected by this gauge mode.
The structure of this eigenvector of the Hessian turns out to be the same for all system sizes we studied.
We expect its eigenvalue to decay exponentially with the system size in the small-$\kappa$ phase.

In the N\'eel phase, $E(\{a_{rr'}\})$ need not be short-ranged and the argument above fails. Indeed, the spinons are charged particles
for the gauge field and their condensation gaps out the gauge degrees of freedom (Anderson-Higgs mechanism).
In the N\'eel phase, computing the boson energy in presence of a global flux (through loops encircling the torus) amounts to impose some twist on the spin directions and the finite spin stiffness $\rho$
naturally leads to an energy cost proportional to the square of the flux, and leads to a finite  Hessian eigenvalue proportional to $\rho$.
In turn, the finite Hessian gap in the N\'eel phase indicates a relative stability of the mean field state with respect to Gaussian $1/N$ corrections.

\subsubsection{Excited mean field solutions for $\kappa=1$.}

\begin{figure}
\includegraphics[width=8cm]{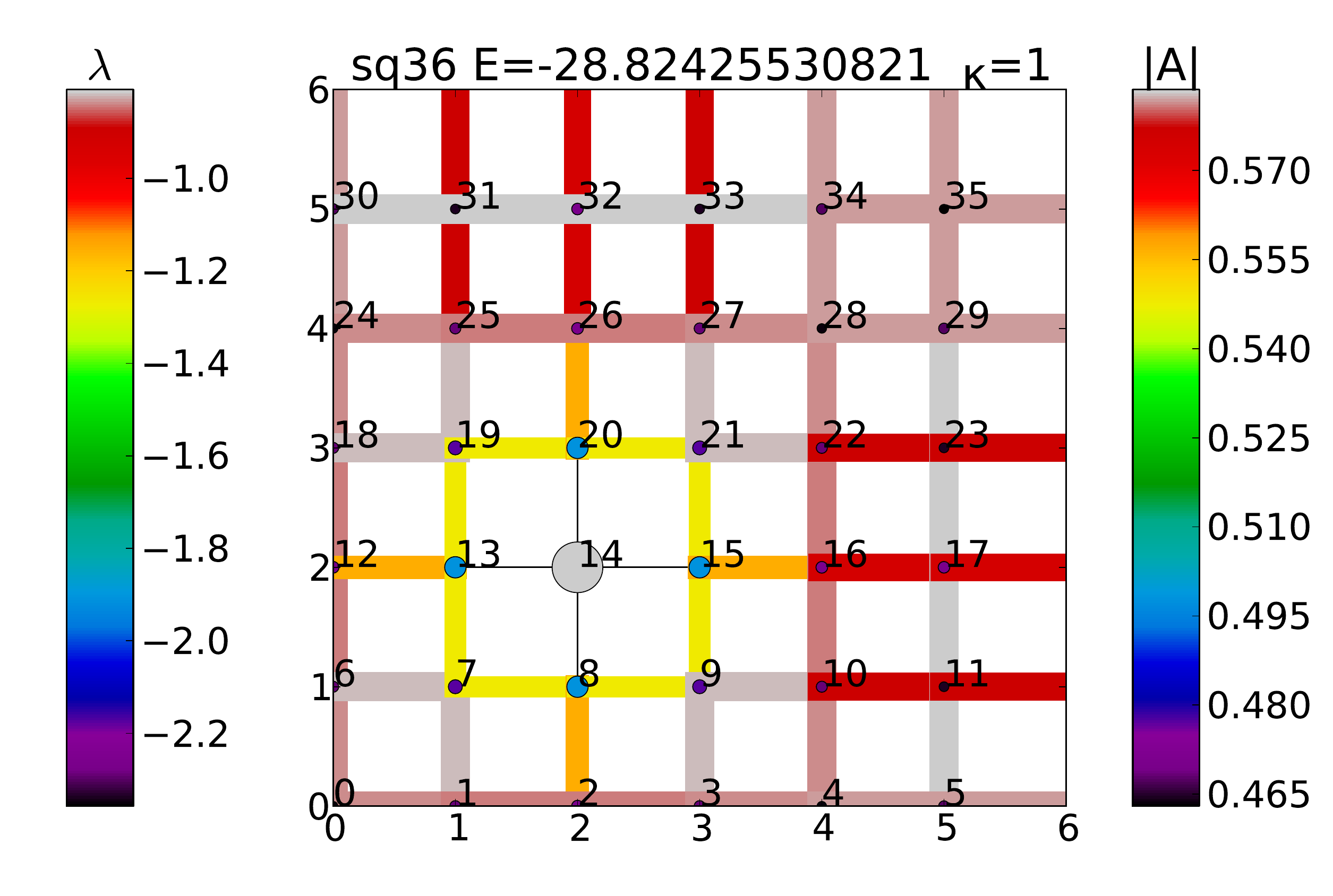}
\caption{(Color online) First excited mean field solution on a square lattice with 36 sites and $\kappa=1$.  The modulus of the bond parameter (right color scale)
is indicated, as well as the chemical potentials (left scale). This solution shows a localized excitation (here around site number 14).
The flux decays rapidly with distance from the center: $0.4046133\pi$ on plaquette [8,9,15,14], $0.037028\pi$ on plaquette $[1,2,8,7]$, $-0.005291300\pi$ on $[0,1,7,6]$, $0.0001131\pi$
on $[4,5,11,10]$, etc. See Fig.~\ref{fig:sq36_e1b_S=0.5} for some the spin-spin correlations in this state.}
\label{fig:sq36_e1_S=0.5}
\end{figure}

For  $\kappa=1$ (N\'eel phase), our search for saddle point on the 36 site cluster shows
that the ground state ($-30.93605205126$) is well separated from the first excited saddle point ($E=-28.82425530821$, Fig.~\ref{fig:sq36_e1_S=0.5}).
No local minimum was found at low energy (see Tab.~\ref{tab:sq36_S=0.5}), but only saddle points.
In addition, the excited saddle points turn out to have very unstable directions (strongly negative Hessian eigenvalues).
There might be some other mean field states in the energy range of Tab.~\ref{tab:sq36_S=0.5}, but these should have rather small basin of attraction with respect to our search algorithms,
since this list of the twelve lowest energy state is stable after thousands of runs starting from random initial conditions. 

\begin{figure}
\includegraphics[width=7cm]{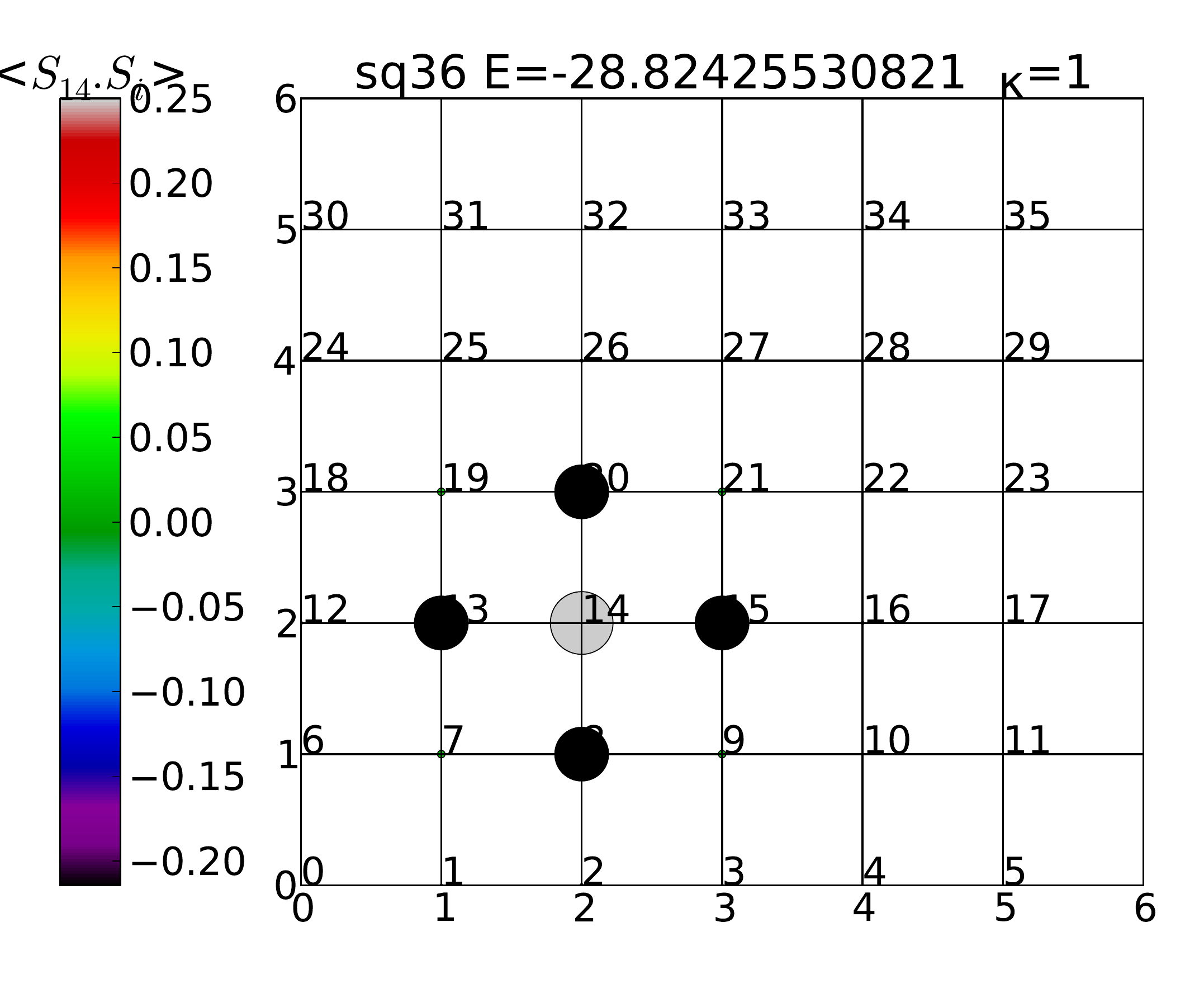}\\
\includegraphics[width=7cm]{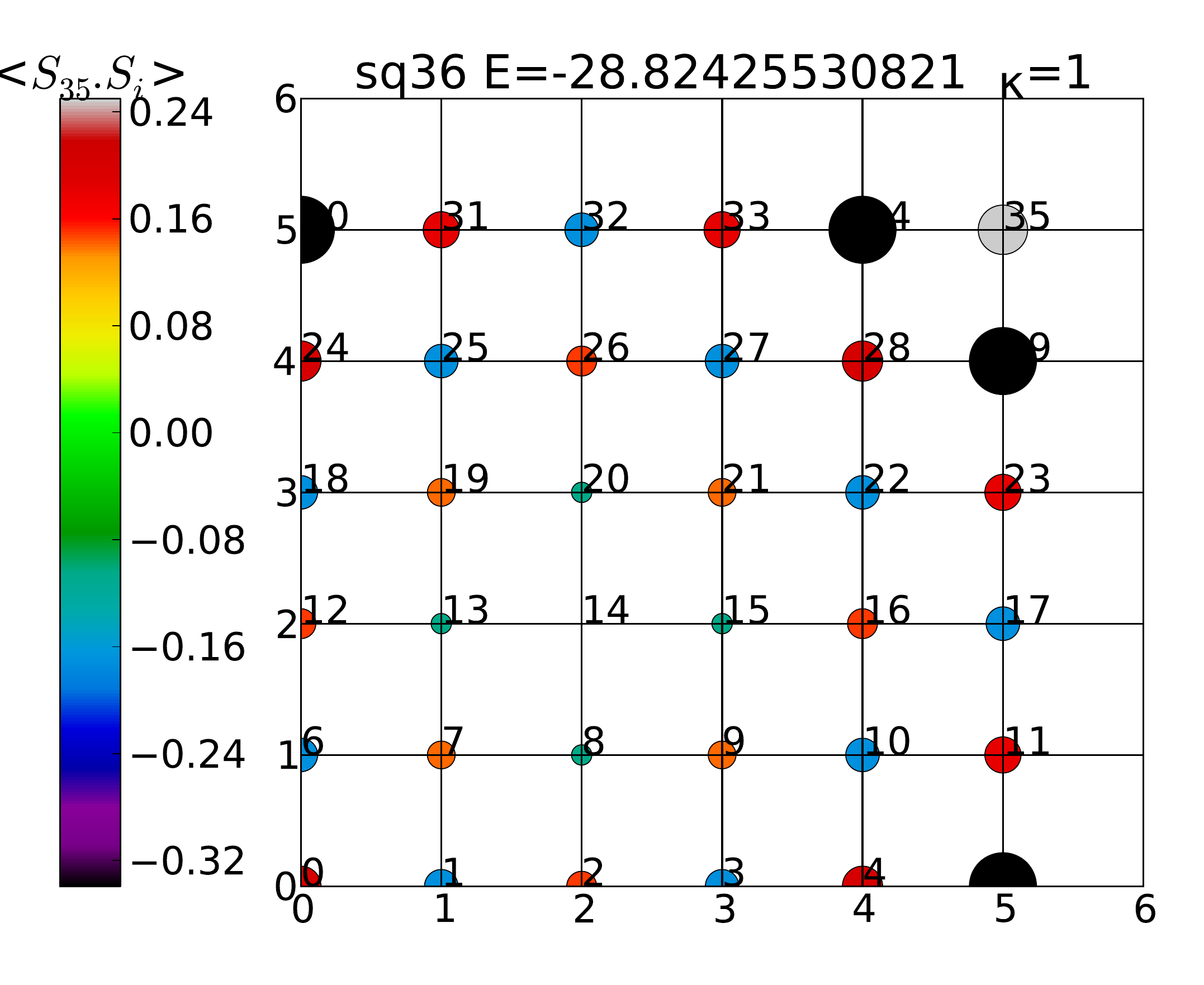}
\caption{(Color online) Spin-Spin correlation in the first excited of a 36-site square lattice with $\kappa=1$ (same state
as in Fig.~\ref{fig:sq36_e1_S=0.5}). The radius of the circle on site $j$  proportional to the correlation $|\langle\vec S_{i0}\vec S _j\rangle|$ with the reference site $i0$
(the sign is available on the color scale).
Top: the reference spin $i_0=14$ is at the center of the localized excitation. Bottom: the reference spin $i_0=35$ is ``far'' from  the center of the  excitation.}
\label{fig:sq36_e1b_S=0.5}
\end{figure}

The first excited saddle point is displayed in Fig.~\ref{fig:sq36_e1_S=0.5}. It takes the form of an excitation localized around one site (here site number 14).
This state is chiral, with a flux equal to $\pm0.40461332\pi$ on the four square plaquettes touching the center of the excitations. The flux then decreases quickly
with distance (see caption of Fig.~\ref{fig:sq36_e1_S=0.5}), but the fact that some non-trivial fluxes (different from 0 or $\pi$) around the center of the excitation shows that
it induces some non-planar spin-spin correlations. Far from the center of the excitation the spins remain in a collinear and in an ordered structure,
as can be seen on the spin-spin correlations displayed in the bottom panel of Fig.~\ref{fig:sq36_e1b_S=0.5}. As for the center spin (number 14), it is correlated with its neighbors,
but it is very weakly correlated ($\langle \vec S_{14} \cdot \vec S_i\rangle\simeq 0$) with the sites which are far apart (top of Fig.~\ref{fig:sq36_e1b_S=0.5}).
Although the system is magnetically ordered, this point-like ``defect'' does not seem do have a simple semi-classical analog.
Interestingly, a similar point-like ``defect'' excitation is found in the ordered phase of
the triangular lattice.

\begingroup
\squeezetable
\begin{table*}
 \begin{tabular}{|c|c|c|c|ccc|ccc|c|}
\hline
$E$ & $\Delta$ & $H$ & d & $N_{\lambda}$ & min$_{\lambda}$ & max$_{\lambda}$  & $N_{|A|}$ & min$_{|A|}$ & min$_{|A|}$ &\\
\hline
-30.93605205126 & 0.14402872021& 2.6295599434& 1& 1& -2.33625413623& -2.33625413623& 1& 0.58295256650& 0.58295256650 &R \\
-28.82425530821$^a$ & 0.12128968489& -7.7009935434& 36& 10& -2.35767326534& -0.80802876988& 12& 0.46297459737& 0.58363641125 &C \\
-28.78219048843 & 0.10001664198& -50.5062824348& 72& 13& -2.36085237706& -0.70467719552& 18& 0.45635222273& 0.59393955633 &R \\
-28.36943538503 & 0.15018056150& -26.9056949522& 72& 12& -2.37180295956& -1.25427516482& 25& 3.06482822239e-15& 0.78318014761 &C \\
-27.87811171341 & 0.10635799592& -48.4908669283& 72& 12& -2.37611640988& -1.20158916706& 18& 0.33406678717& 0.62744479349 &R \\
-27.67404431445$^b$ & 0.24030790672& -33.3227937772& 2& 1& -2.12079971339& -2.12079971339& 2& 0.52769740399& 0.59657118301 &R \\
-27.58542591653 & 0.12018732985& -11.9733712946& 72& 16& -2.38843405218& -1.28794576300& 24& 0.29836817056& 0.63116481780 &C \\
-27.55178404956 & 0.08439099361& -67.0253600361& 72& 16& -2.48677863782& -1.13200578463& 24& 3.38069859599e-16& 0.64685457030 &C \\
-27.52461818853 & 0.19788831856& -3.8580290414& 36& 6& -2.23233357411& -1.87795360324& 14& 0.45547996236& 0.59859663572 &C \\
-27.49457258263 & 0.18888774623& -1.9063958450& 36& 8& -2.24866064506& -1.74743986196& 14& 0.46141122781& 0.60087012841 &C \\
-27.49456554040 & 0.19367584933& -5.7251545034& 36& 6& -2.22476689000& -1.75751056646& 14& 0.34656259228& 0.60092322028 &C \\
-27.40385237877 & 0.16760744308& -21.1217748777& 144& 18& -2.32248736876& -1.30042949863& 38& 0.27024089067& 0.64068344661 &C \\
\hline
\end{tabular}
\caption{Low-energy saddle points for a square lattice cluster with 36 sites and $\kappa=1$.
From the left column to the right: energy, spinon gap, lowest non-zero Hessian eigenvalue, degeneracy, number of different chemical potentials $\lambda$ (1 means uniform, etc;), minimum and maximum values of $\lambda$, number of different values of $|A|$
minimum and maximum values of $|A|$, real (R) or complex/chiral solution (C).
The spatial modulations
of $|A_{ij}|$ and $\lambda_i$ for solution $^a$ are displayed in Fig.~\ref{fig:sq36_e1_S=0.5}. Solution $^b$ has
the same spatial structure (two-fold degeneracy) as the one shown in Fig.~\ref{fig:sq36VBC}b.}
\label{tab:sq36_S=0.5}
\end{table*}
\endgroup

\subsubsection{Excited mean field solutions for $\kappa=0.1$.}

For $\kappa=0.1$, the ground state on the square lattice is still uniform and without any flux, but the spinon gap (Fig.~\ref{fig:gap_hess_sq}) remains finite in the thermodynamic limit.
This mean field state has been argued\cite{rs89} to be unstable  at finite $N$, due to strong gauge fluctuations.
The finite $N$ ground state is believed to spontaneously break some lattice symmetry to form a valence-bond crystal (VBC).
In fact, the mean field energy landscape is very different from the one observed in the the magnetic phase.
The first local minima and saddle points are listed in Tab.~\ref{tab:sq36_S=0.05}.

\begingroup
\squeezetable
\begin{table*}
 \begin{tabular}{|c|c|c|c|ccc|ccc|c|}
\hline
$E$ & $\Delta$ & $H$ & d & $N_{\lambda}$ & min$_{\lambda}$ & max$_{\lambda}$  & $N_{|A|}$ & min$_{|A|}$ & min$_{|A|}$ &\\
\hline
-1.83492169860 & 0.40562447718& 0.0514862091& 1& 1& -0.62322660824& -0.62322660824& 1& 0.11828994799& 0.11828994799 &R \\
-1.83349695884$^a$ & 0.43766690295& -0.0513757991& 2& 1& -0.62174106995& -0.62174106995& 2& 0.11707639293& 0.11940834877 &R \\
-1.83265972180 & 0.42301616405& 3.3891521814e-05& 36& 6& -0.62283957972& -0.61901747505& 12& 0.11749362973& 0.11895083747 &C \\
-1.83265969367 & 0.42301757051& -3.3879532685e-05& 72& 12& -0.62289751868& -0.61872552844& 25& 0.11742978176& 0.11900687718 &C \\
-1.83265966553 & 0.42301846049& -3.3877477372e-05& 36& 10& -0.62295399258& -0.61842513968& 12& 0.11745616476& 0.11898150831 &C \\
-1.83162471056 & 0.41456789451& -0.2513880759& 36& 6& -0.62576828556& -0.60318829800& 12& 0.10370572327& 0.13006171701 &C \\
-1.83121683711 & 0.41375814481& -0.2872095650& 72& 12& -0.63567288324& -0.58931901135& 25& 0.03077944153& 0.13945180759 &C \\
-1.83088353771 & 0.43032855876& -0.2737625727& 72& 9& -0.62660943591& -0.60282776054& 24& 0.10017310034& 0.13081740450 &C \\
-1.83047599540 & 0.42976957584& -0.3191268237& 72& 12& -0.63670770779& -0.58921985280& 25& 0.01544819043& 0.13979752631 &C \\
-1.83036799495 & 0.43107796748& -0.3414892108& 72& 12& -0.63601360482& -0.58745590945& 25& 0.00522910878& 0.14104278833 &C \\
-1.83004203619 & 0.43897814318& -0.2774384716& 36& 6& -0.62530891168& -0.60252710044& 12& 0.10277137746& 0.13077306110 &C \\
-1.82976553423$^c$ & 0.44236450064& 0.0004197766& 9& 3& -0.62142264885& -0.61823831215& 4& 0.11796026878& 0.11822830533 &C \\
$\cdots$&&&&&&&&&&\\
-1.82534862050$^b$ & 0.45598298556& 0.0049547827& 4& 1& -0.61651000360& -0.61651000360& 2& 0.11510636037& 0.12084117665 &C \\
\hline
\end{tabular}
\caption{Low-energy saddle points for a square lattice cluster with 36 sites.
$\kappa=0.1$. Solutions $^a$ and $^b$ are displayed in Fig.~\ref{fig:sq36VBC}, and solution $^c$ is shown in Fig.~\ref{fig:sq36VBC2}.}
\label{tab:sq36_S=0.05}
\end{table*}
\endgroup

\begin{figure}
\includegraphics[width=8cm]{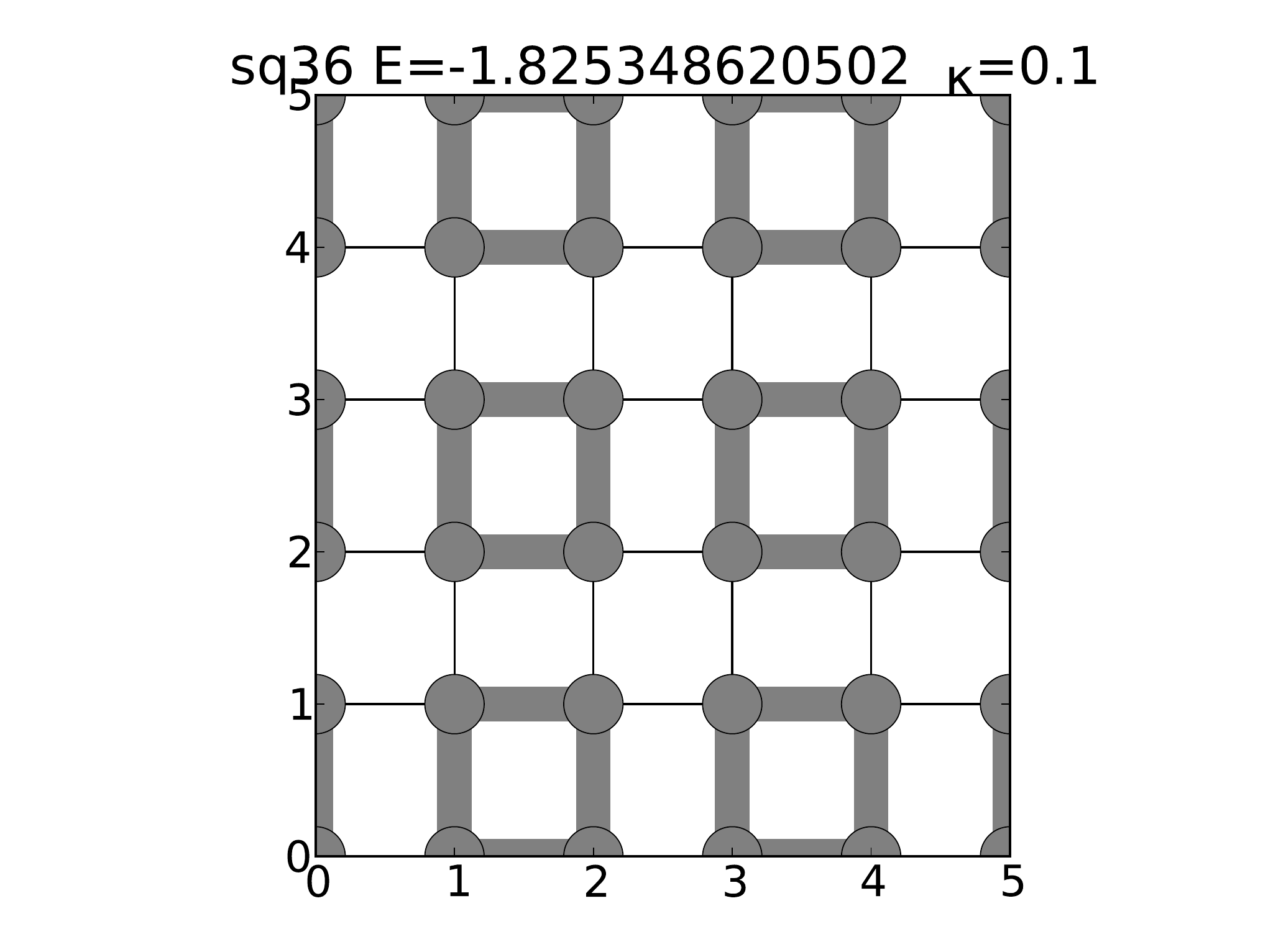}
\includegraphics[width=8cm]{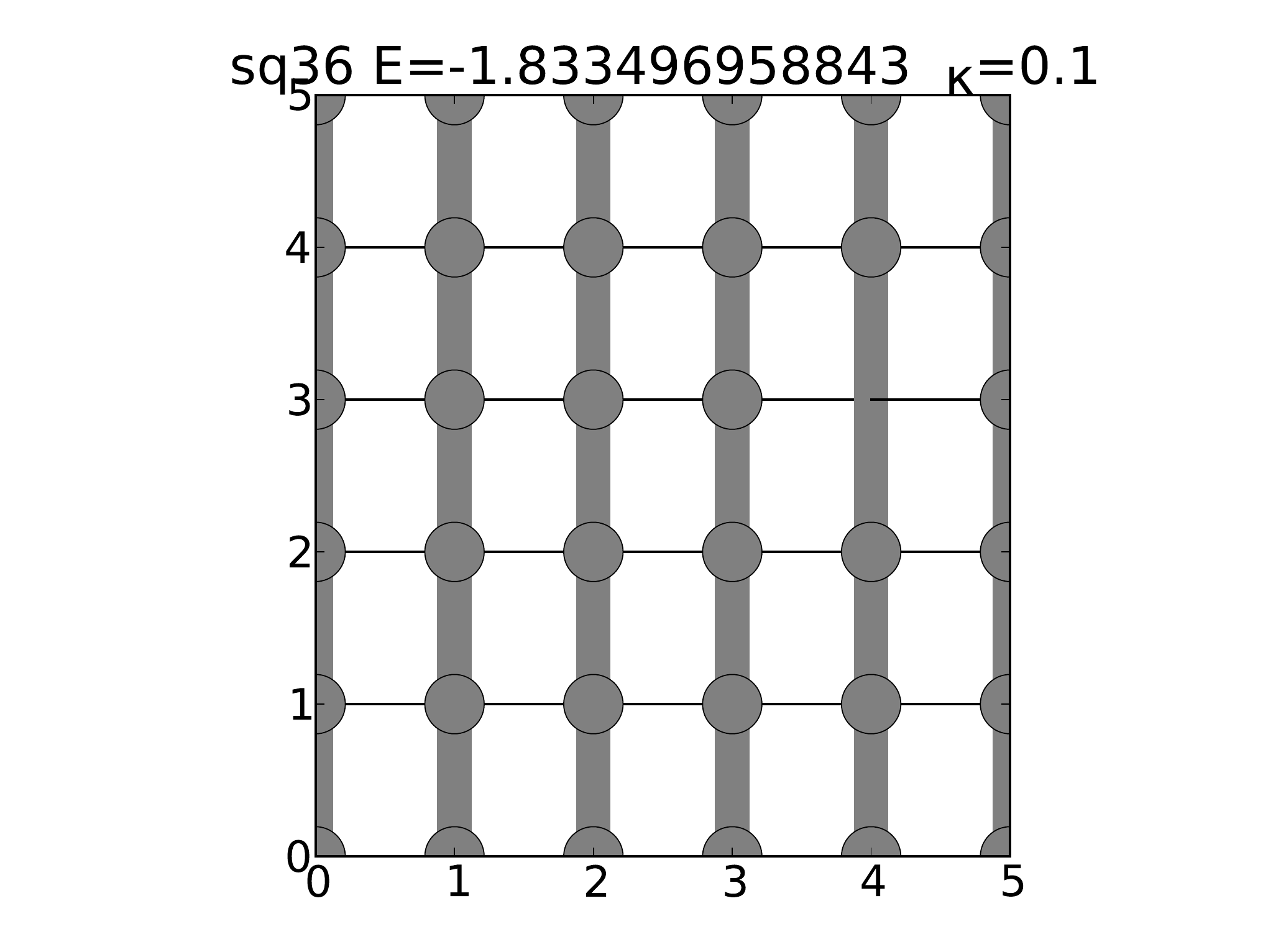}
\caption{Two low-energy mean field solutions on a square lattice with 36 sites and $\kappa=0.1$.  $|A|$ takes only two values and the bonds with the stronger $|A|$ are
indicated by fat grey lines (for values, see Tab.~\ref{tab:sq36_S=0.05}). All the sites are equivalent and the chemical potential is uniform in both cases.}
\label{fig:sq36VBC}
\end{figure}

\begin{figure}
\includegraphics[width=9cm]{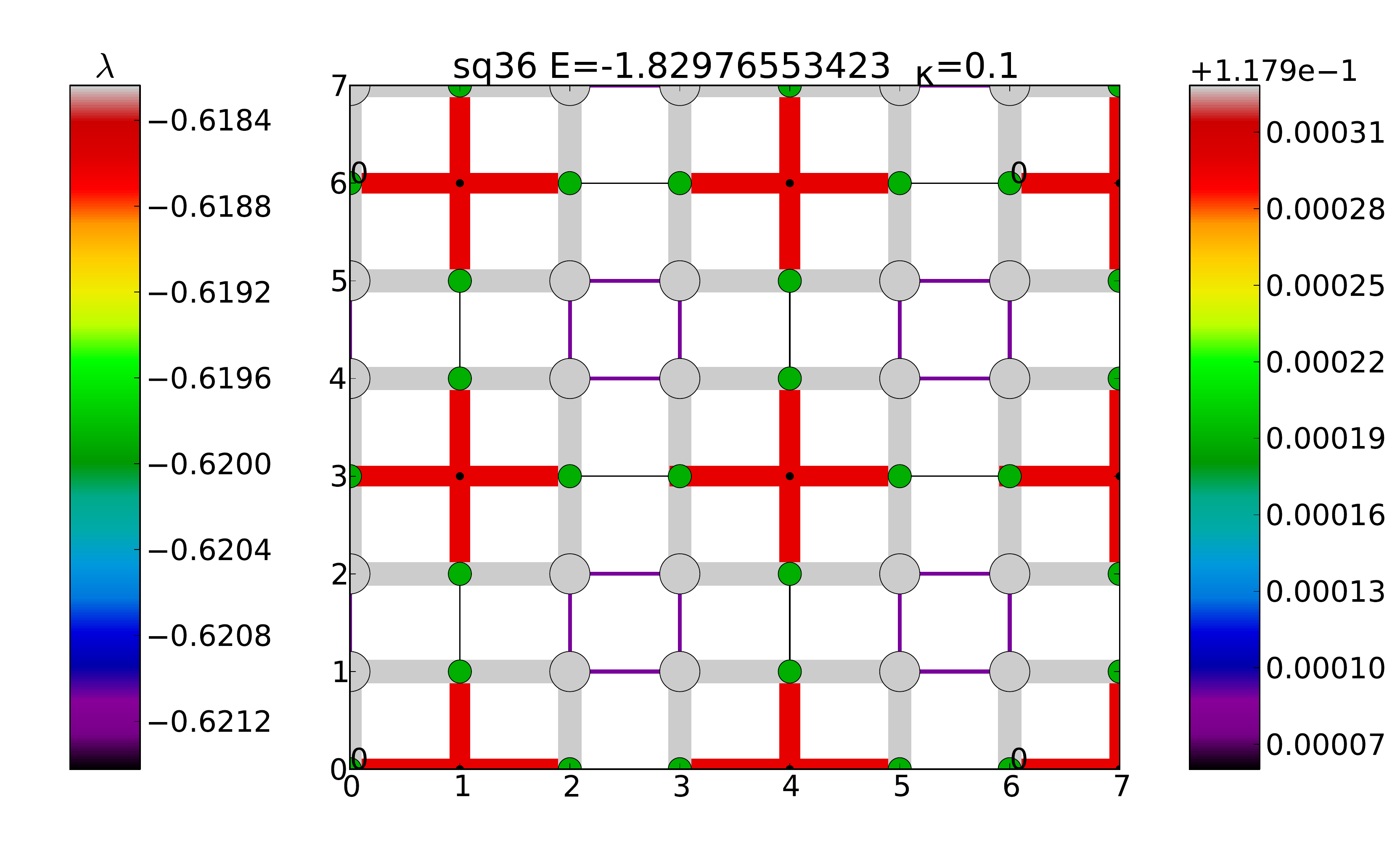}
\caption{(Color online) A low-energy mean field solution on a square lattice with 36 sites and $\kappa=0.1$.  The modulus of the bond parameters is indicated -- the right color scale shows deviations from the minimum value.
The left color scale indicates the chemical potentials (dot on each site). This solution shows modulations with the symmetries of a $3\time3$ unit cell VBC.}
\label{fig:sq36VBC2}
\end{figure}

We observed a high density of local minima and saddle points, with very small Hessian eigenvalues.
This high density of excited saddle points  reflects the presence of some strong gauge fluctuations at (large but) finite $N$.
We note that several such saddle points display modulations of the bond amplitudes $|A|$ which have the same symmetries as
the VBC previously considered for square lattice antiferromagnets.
The first excitations (line $^a$ in Tab.~\ref{tab:sq36_S=0.05}) shows a one-dimensional modulation, which is predicted to occur at finite $N$ when $\kappa=2\;{\rm mod}\; 4$.\cite{rs89} 
And among the highly symmetric solutions we also note a plaquette VBC ($E=-1.82534862050$) (Fig.~\ref{fig:sq36VBC}).
This state does however not correspond to a simple VBC since it is a complex/chiral solution with non-trivial (different from 0 or $\pi$) fluxes on all the square plaquettes.
See also Fig.~\ref{fig:sq36VBC2} for another VBC-like state.
According to the analysis of Read and Sachdev, non-perturbative gauge fluctuations (proliferation of hedgehogs point-like instantons) are responsible for the lattice symmetry breaking
at finite $N$, leading to a modulation of $|A|$ which is exponentially small in $N$.
The mechanism is somewhat different here since we observe VBC-like low-energy saddle points although gauge fluctuations are completely absent (all the $A_{ij}$ are frozen in the SBMFT).

\subsection{Triangular lattice}

\subsubsection{Hessian of the ground state}

\begin{figure}
\includegraphics[angle=-90,width=7cm]{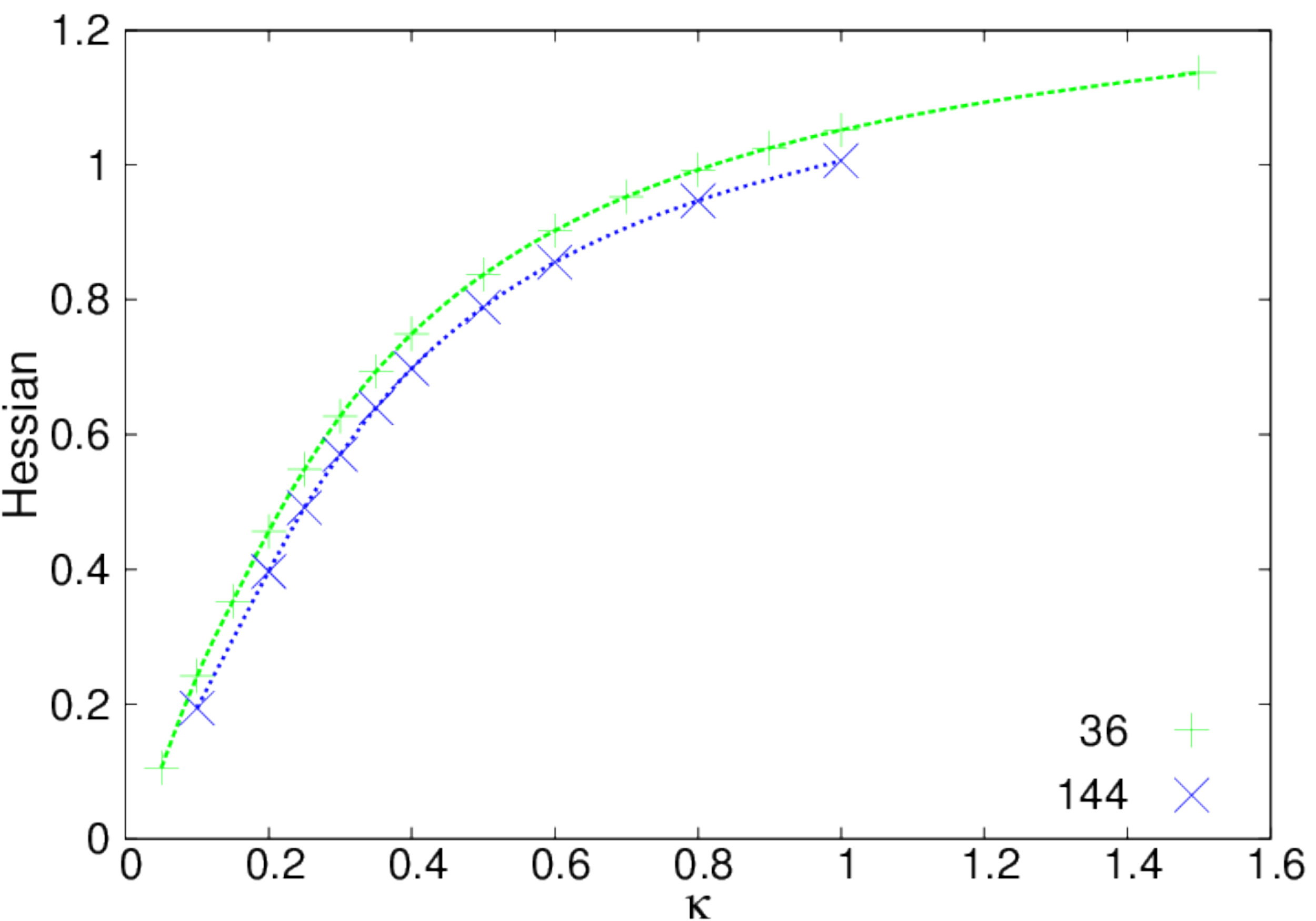}
\includegraphics[angle=-90,width=7cm]{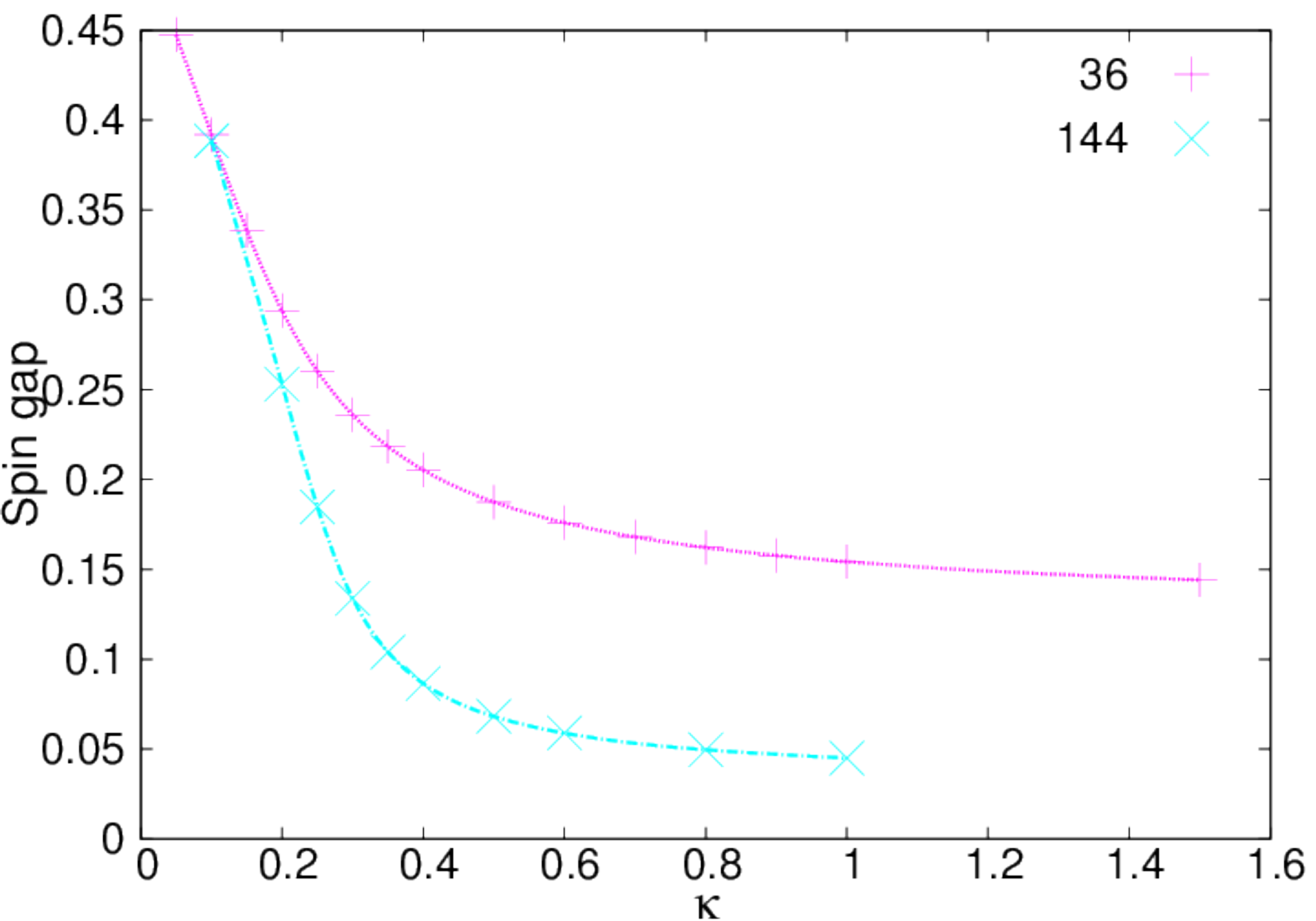}
\caption{(Color online) Top panel: smallest eigenvalue of the Hessian for the ground state state on 36-site and 144-site triangular clusters. Bottom: spin gap.
In the thermodynamic limit this state is associated to magnetic long-range order (vanishing spin gap) for $\kappa\gtrsim 0.34$.\cite{sachdev92}
}
\label{fig:gap_hess_tri}
\end{figure}

On the triangular lattice we find that the lowest energy state corresponds to the (spatially uniform) solution
studied by Sachdev\cite{sachdev92} (0-flux state~\cite{wv06}) and  leads to magnetic long-range order ($\sqrt{3}\times\sqrt{3}$)
for $\kappa>0.34$, and a gapped and deconfined $\mathbb{Z}_2$ liquid for  $\kappa\lesssim0.34$.
As usual, these two phases can be distinguished by the spinon gap: it drops to zero when increasing the system size in the magnetic phase and stays finite in the
liquid phase (see Fig.~\ref{fig:gap_hess_tri}).

The Hessian has a large lowest eigenvalue, which indicates the stability of this mean field state with respect to $1/N$ fluctuations.
The evolution of this lowest eigenvalue is plotted in
Fig.~\ref{fig:gap_hess_tri}.  It slightly decreases with the system size (comparing 36 and 144 sites), but is certainly finite in the thermodynamic limit.
Contrary to the square lattice situation, one does not detect any dramatic change of behavior between the gapped  phase and the magnetic one.

In fact, since the lattice is no bipartite, the IGG of the uniform mean field state is discrete ($\mathbb Z_2$) and we do not expect any gapless
modes associated to  {\it small} perturbations $A_{ij}=A_{ij}^0+dA_{ij}=$ with $ |dA_{ij}|\ll1$. Still, important $\mathbb Z_2$ (gapped) gauge excitations are expected in the spin liquid phase, and they
will be discussed in Sec.~\ref{ssec:visons}. 

\begin{figure}
\hspace*{-2.2cm}\includegraphics[width=11cm]{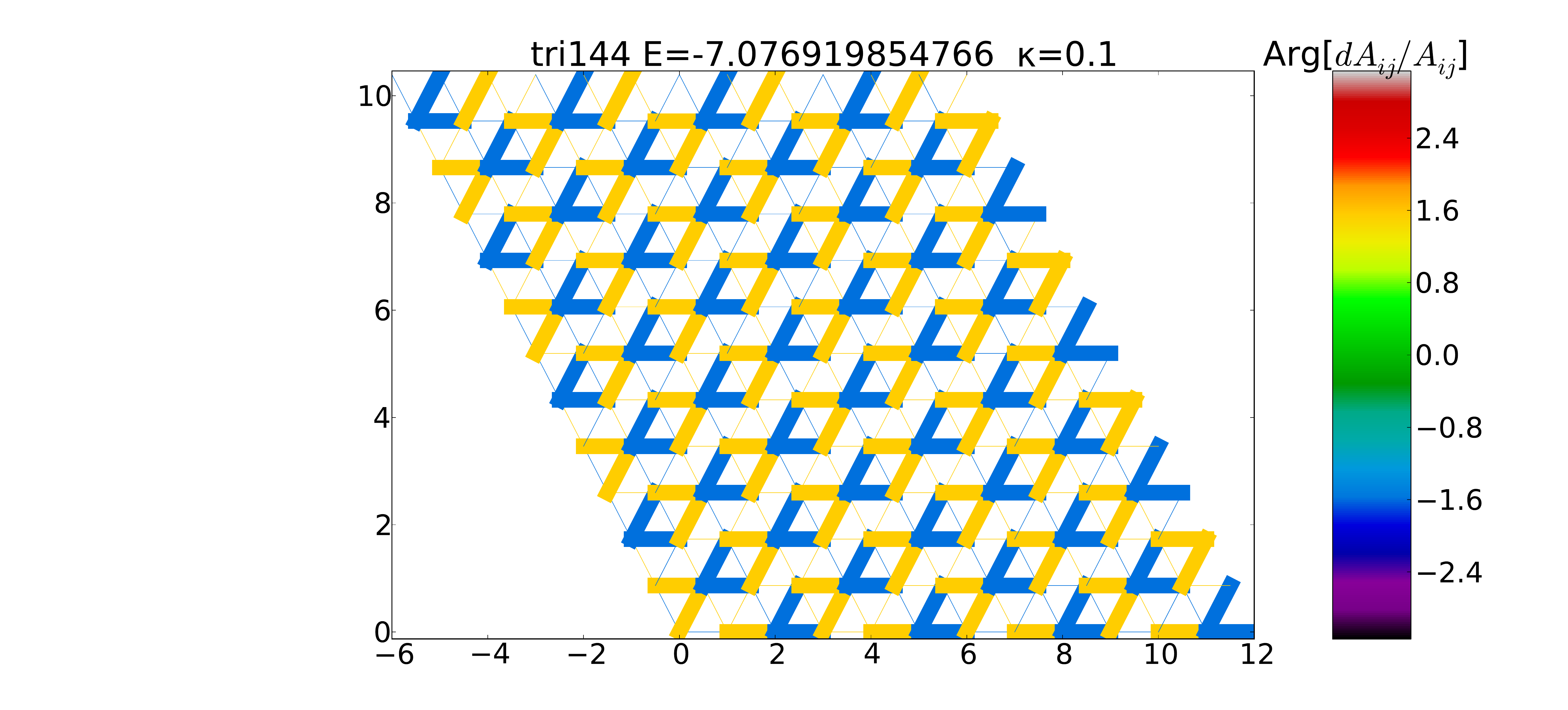}\\
\hspace*{-2.2cm}\includegraphics[width=11cm]{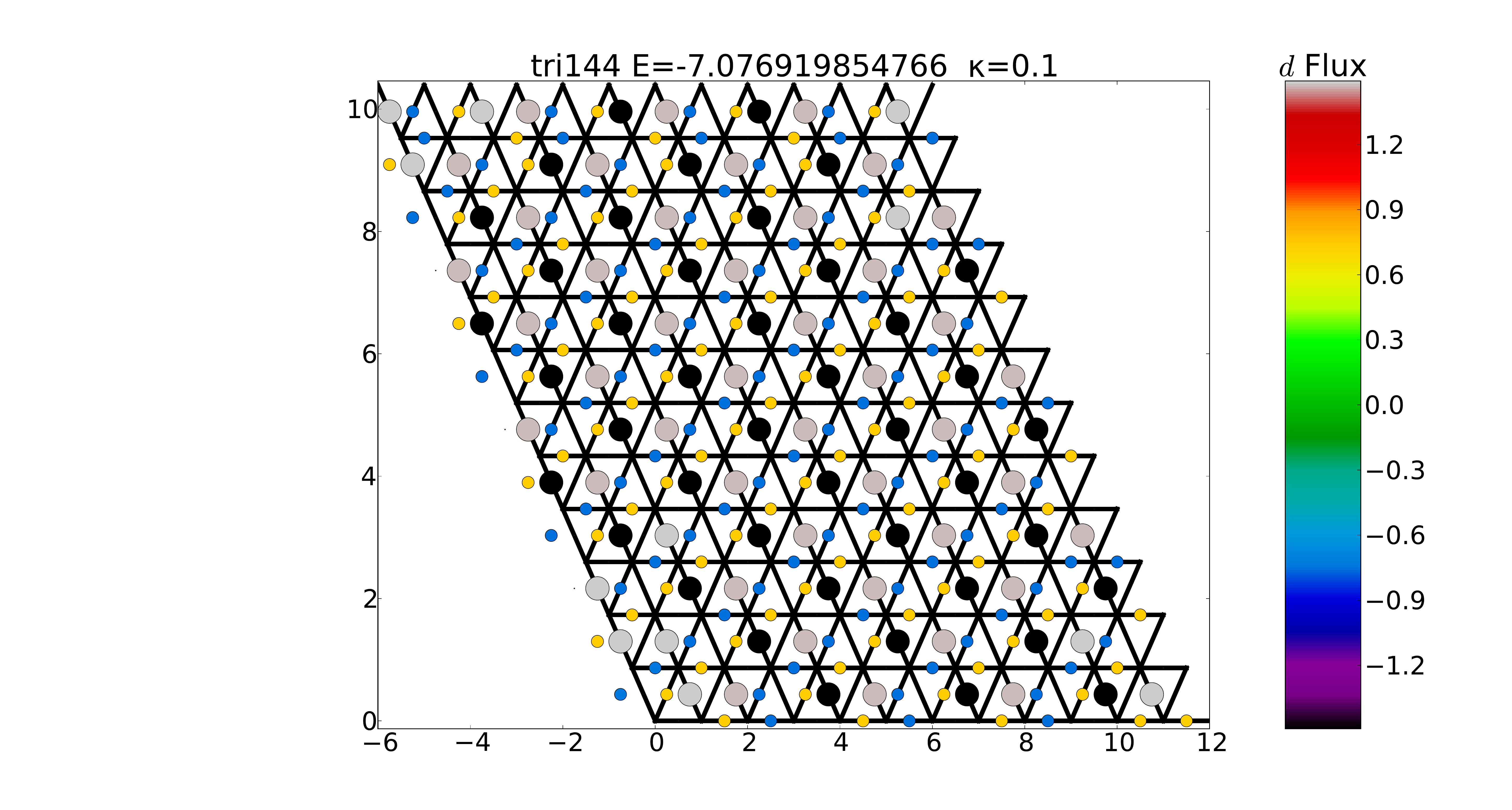}
\caption{(Color online) Hessian eigenvector $dA/A$ corresponding to the smallest (degenerate) eigenvalue (0.1943) for a 144-site triangular lattice with $\kappa=0.1$ (uniform ground state).
The complex argument of $dA/A$ take only two values: $\pm\pi/2$ (blue and yellow), indicating that $dA$ is a (gapped U(1)) gauge excitation.
Bottom: Infinitesimal flux variation $dF$ associated to the gauge mode above. For each diamond the magnitude of the flux variation is indicated by the radius
of the dot in its center (see color scale for the sign).}
\label{fig:tri144_hev}
\end{figure}

The lowest eigenvector of the Hessian is represented in Fig.~\ref{fig:tri144_hev} for  $\kappa=0.1
$. As in all the cases we looked at, it correspond to a {\it phase} fluctuation
of the bond variables. The associated flux modulations for all diamond loops are shown in the bottom panel of the figure (note that the sign of each flux is somehow arbitrary since it depends on the choice of
an origin of the loop). This mode represent the lowest energy U(1) gauge excitation. It is gapped  since $U(1)$ is {\it not} the low-energy gauge group (IGG)
of this mean field state.

\subsubsection{Excited mean field solutions}

\begin{figure}
\includegraphics[width=9cm]{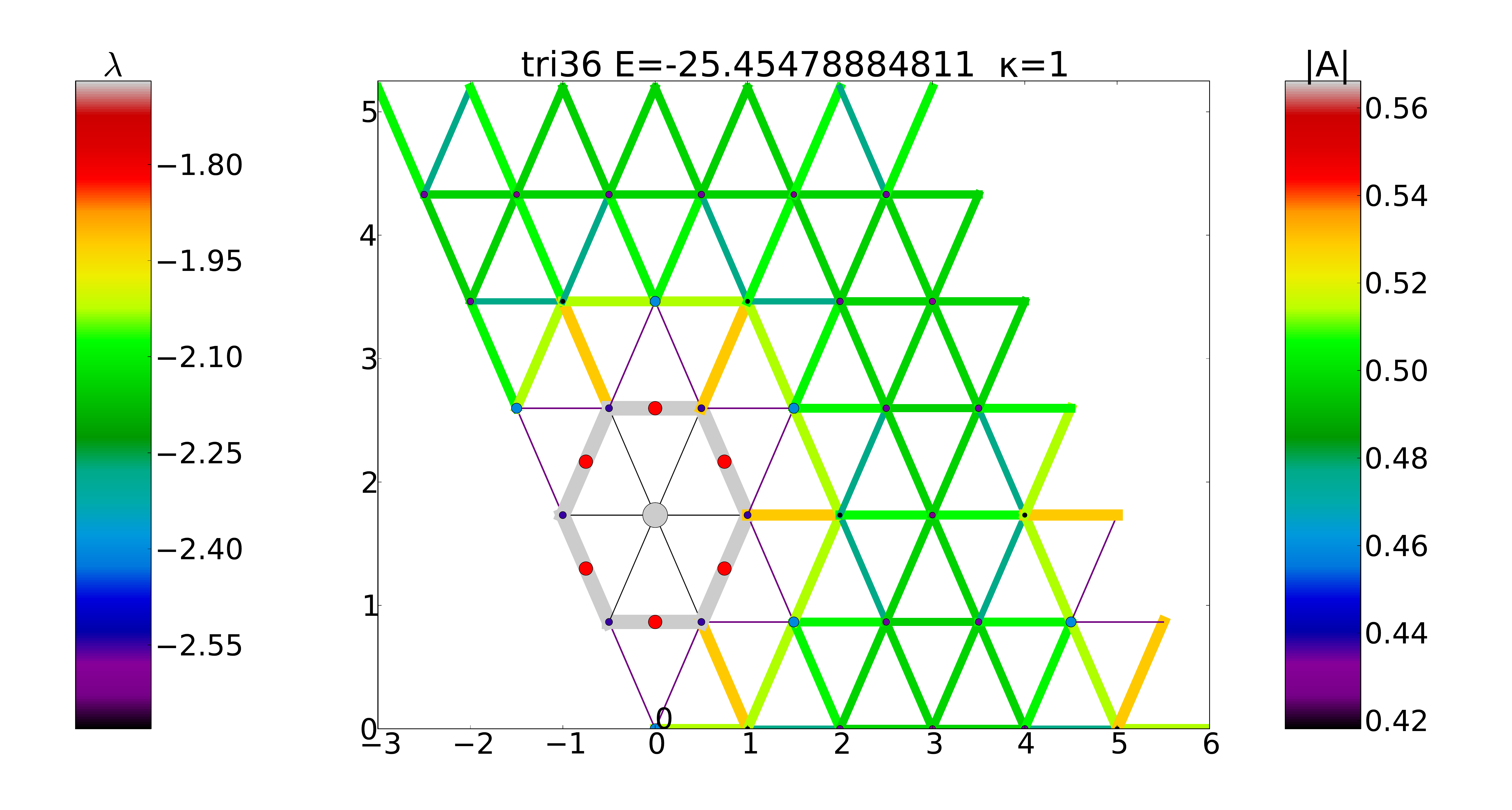}
\caption{(Color online) First excited state on a triangular cluster with 36 sites and $\kappa=1$. The bond moduli are invariant by lattice rotations about the site largest chemical potential (grey circle).
The flux vanish on all diamonds except for the six diamonds which diagonal bond is marked by a red dot. The later have flux $\pi$.
Notice the similarity with Fig.~\ref{fig:sq36_e1_S=0.5}.
}
\label{fig:tri36b}
\end{figure}

In Tab.~\ref{tab:tri36_S=0.5} we list these first mean field states obtained for $\kappa=1$ in a 36-site sample.
As for the square lattice in its magnetic phase, we only find saddle points and no local minima among the first states.
The first excited saddle point
is at an energy $0.94927$ above the ground state and this gap is very likely finite in the thermodynamic limit.
This excited state has the spatial structure of a localized excitations (Fig.~\ref{fig:tri36b}) which  resembles that of the square lattice.
Finally, we note that contrary to the square lattice case, the lowest energy states of Tab.~\ref{tab:tri36_S=0.5} have  real bond amplitudes
(more precisely: can be made real with an appropriate gauge choice),
indicating the coplanar nature of their spin-spin correlations.

\begingroup
\squeezetable
\begin{table*}
 \begin{tabular}{|c|c|c|c|ccc|ccc|c|}
\hline
$E$ & $\Delta$ & $H$ & d & $N_{\lambda}$ & min$_{\lambda}$ & max$_{\lambda}$  & $N_{|A|}$ & min$_{|A|}$ & min$_{|A|}$ &\\
\hline
-26.40405992819 & 0.15424645309& 1.0517831524& 1& 1& -2.58830049633& -2.58830049633& 1& 0.49723336391& 0.49723336391 &R \\
-25.45478884810 & 0.14039937369& -2.3583827803& 36& 7& -2.68083518864& -1.66967911414& 12& 0.41812686812& 0.56615895080 &R \\
-24.89782095003 & 0.13788292075& -2.1218542408& 108& 12& -2.66775663053& -1.86004294873& 31& 0.25950830810& 0.56907182479 &R \\
-24.88367410179 & 0.11975325026& -2.3419680753& 432& 36& -2.76307967976& -1.75626091358& 108& 0.13936678359& 0.57983094362 &R \\
-24.88248919944 & 0.14303067790& -2.3536310845& 216& 18& -2.72365738218& -1.77512884614& 56& 0.18149763138& 0.57925175603 &R \\
-24.85225639276 & 0.05506495561& -2.6494385700& 216& 21& -2.87033693396& -1.82953713470& 57& 0.05089520618& 0.56602789699 &R \\
-24.54012950533 & 0.07534371832& -2.7179265074& 216& 18& -2.83154574151& -1.61720470431& 56& 0.25139432620& 0.57460696668 &R \\
-24.52557412388 & 0.06458137688& -2.4327487996& 216& 18& -2.82124110423& -1.87789249232& 56& 0.09674702612& 0.56912072760 &R \\
-24.51689402152 & 0.10842201321& -2.4288970581& 72& 10& -2.79081268614& -1.66076773241& 21& 0.35681606435& 0.56617071630 &R \\
-24.42913380032 & 0.13143243020& -2.5993027532& 54& 8& -2.66560684707& -1.62001869447& 17& 0.38239562932& 0.56552771557 &R \\
-24.38386557808 & 0.12833408469& -2.0204356396& 108& 13& -2.73034611703& -1.88206681620& 30& 0.25362349993& 0.56892183451 &R \\
-24.36651624089 & 0.11045933281& -2.2168211792& 432& 36& -2.76253254084& -1.77578363386& 108& 0.10590160533& 0.58136368633 &R \\
\hline
\end{tabular}
\caption{Energy minimum and low-energy saddle points for a triangular lattice cluster with 36 sites and $\kappa=1$. Notice that the ground state is very stable (large Hessian gap: 1.051) and
that all the other saddle points are unstable (negative Hessian eigenvalue). The gap $-25.454+26.404=0.949$ is quite large.
The ground state ($E=-26.40405992819$) is a solution with vanishing flux on all the diamonds, and three-sublattice long-range spin-spin correlations (the critical value
for magnetic long-range order is $\kappa=0.34$.\cite{sachdev92}).
The bond strength of the first excited state are displayed in Fig.~\ref{fig:tri36b}.
}
\label{tab:tri36_S=0.5}
\end{table*}
\endgroup

\begin{figure}
\hspace*{-2cm}\includegraphics[width=10cm]{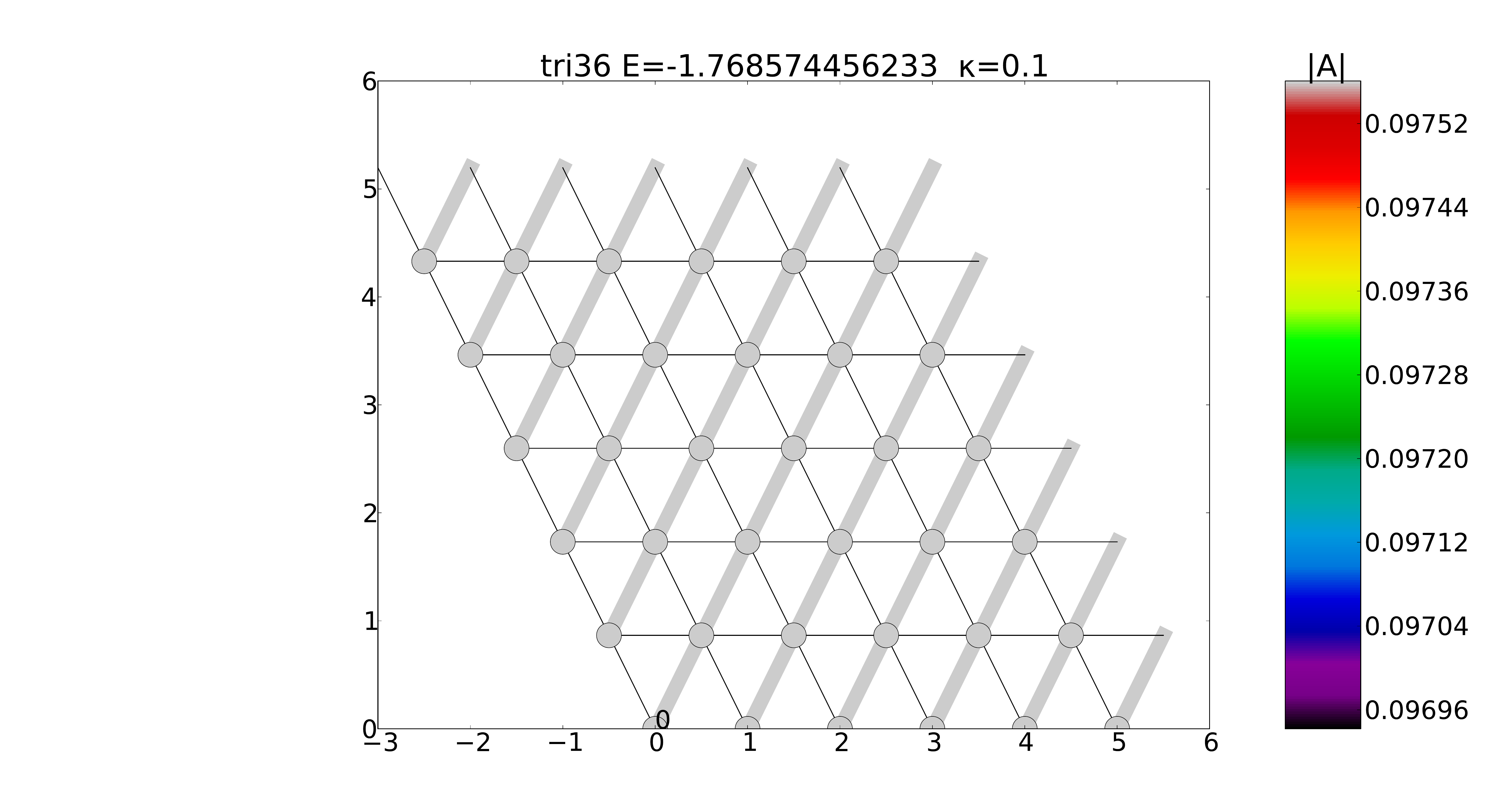}
\caption{First excitation on a triangular cluster with 36 sites and $\kappa=0.1$. This state is three-fold degenerate. With the ground state, it forms  the four-fold topological degeneracy expected
for a $\mathbb{Z}_2$ liquid on a torus. Notice (scale on the right) that the difference between the largest $|A_{ij}|$  and the smallest one is only $\sim 6.10^{-4}$ and should vanish in the thermodynamic limit.
}\label{fig:tri36_S=0.05b}
\end{figure}

For small enough $\kappa$, the SBMFT describes a gapped spin liquid of $\mathbb Z_2$ type.\cite{sachdev92,wv06}
Tab.~\ref{tab:tri36_S=0.05} gives the first saddle points obtained at $\kappa=0.1$. The ground state is uniform and all the rhombi have a vanishing flux, as expected.
The first excited saddle point has slight modulations ($\sim 6.10^{-4}$) of the bond amplitudes (Fig.~\ref{fig:tri36_S=0.05b}). This state also has vanishing fluxes
on all the rhombi, but it differ from the ground state
by the presence of an  additional flux $\pi$ along some long loop winding around the torus (three possible choices).
These three states become homogeneous in the thermodynamic limit  and degenerate with the ground state. With the ground state
they form  the four-fold topological degeneracy of the $\mathbb{Z}_2$ liquid on a torus.\cite{rc89,rs91}
These states were found by the optimization algorithm starting from random initial conditions, and not ``forced'' by hand.
We are thus confident that the method is able to find the low energy solutions in a systematic way on small clusters at least up to a few tens of sites and bonds.

Above the four topological ground states we observe many saddle points which do not
show any simple/regular spatial pattern (high number of inequivalent sites and bonds). These states have energies
significantly below the first vison-pair state we have found (last line in Tab.~\ref{tab:tri36_S=0.05}
and Fig.~\ref{fig:visons_tri36}). Due to the presence of complex fluxes (not $0$ or $\pi$), these states do not have
a simple interpretation in terms of the $\mathbb{Z}_2$ gauge field.
The presence of these some additional degrees of freedom  is somewhat intriguing since the  low-energy description of such a short-ranged RVB phase 
is expected to be $\mathbb{Z}_2$ a gauge theory coupled to gapped spinons.

\begingroup
\squeezetable
\begin{table*}
 \begin{tabular}{|c|c|c|c|ccc|ccc|c|}
\hline
$E$ & $\Delta$ & $H$ & d & $N_{\lambda}$ & min$_{\lambda}$ & max$_{\lambda}$  & $N_{|A|}$ & min$_{|A|}$ & max$_{|A|}$ &\\
\hline
-1.77115513853 & 0.39183413870& 0.2418792448& 1& 1& -0.63927958247& -0.63927958247& 1& 0.09721004221& 0.09721004221 & R\\
-1.76857445623 & 0.43969994315& 0.1404951036& 3& 1& -0.63620536809& -0.63620536809& 2& 0.09694182615& 0.09756074413 & R\\
-1.76558449027 & 0.40008559551& -0.0136673394& 108& 13& -0.64036695527& -0.61163933625& 30& 0.08631046645& 0.10220320383 & C\\
-1.76556438644 & 0.39881698544& -0.0662494156& 216& 21& -0.64094448900& -0.60807754792& 57& 0.07142251452& 0.11045947514 & C\\
-1.76519815707 & 0.40323866935& -0.0476017475& 216& 21& -0.64101212209& -0.62149883058& 57& 0.08166138572& 0.10941122095 & C\\
-1.76517640350 & 0.40348584568& -0.0574617718& 432& 36& -0.64242077150& -0.62028199544& 108& 0.08102625728& 0.11240116020 & C\\
-1.76512811482 & 0.40255745524& -0.0675887556& 216& 21& -0.64418307267& -0.62192577975& 57& 0.07214207918& 0.11553853523 & C\\
-1.76498033616 & 0.40426965800& -0.0822323880& 216& 18& -0.64027903292& -0.62235496321& 56& 0.07818420245& 0.11549233664 & C\\
-1.76494386569 & 0.40647240030& -0.0077675537& 108& 12& -0.64031335557& -0.61976563929& 31& 0.08656330909& 0.10201782841 & C\\
-1.76490270376 & 0.41103320490& 0.0552851450& 216& 18& -0.64030870204& -0.62463756694& 56& 0.08904499619& 0.10413653140 & C\\
-1.76486322625 & 0.40885376933& -0.0517770911& 432& 36& -0.64144886953& -0.62152076605& 108& 0.08370457723& 0.10627245220 & C\\
-1.76481070231 & 0.40714839383& -0.0588744760& 216& 18& -0.64059451058& -0.62076171059& 56& 0.08285231897& 0.10673538608 & C\\
$\cdots$&&&&&&&&&&\\
-1.76267094982 & 0.41119320956& -0.2136294440& 108& 12& -0.64647502600& -0.60434357136& 31& 0.07784772351& 0.11384461169 & R\\
\hline
\end{tabular}
\caption{Low-energy saddle points for a triangular lattice cluster with 36 sites and $\kappa=0.1$.
The ground state and the first excited saddle point ($E=-1.76857445623$, three fold degenerate)
form the four-fold topological degeneracy.\cite{rc89,rs91} The other saddle points listed here are chiral(complex), except for the last line, which corresponds to a pair of visons (see Sec.~\ref{ssec:visons} and Fig.~\ref{fig:visons_tri36}).}
\label{tab:tri36_S=0.05}
\end{table*}
\endgroup

\subsubsection{Visons}
\label{ssec:visons}

A $\mathbb{Z}_2$ liquid possesses non-magnetic excitations named {\it visons}, which correspond to $\pi$-flux quanta of the effective Ising gauge theory.\cite{rc89,rs89}
In a finite system (without boundaries) the number of vison is necessarily even, and a trial vison-pair state can be constructed as follows.
One starts from the uniform mean field ground state and one reverses the sign of the bond parameters $A_{ij}$ for all the bonds $ij$
crossing a cut extending from the a first plaquette to a second one. The gauge flux is then concentrated in the immediate vicinity of these two plaquettes which correspond to the vison
core positions. Any gauge-invariant operator far from the vison cores  is unaffected by this modification.

\begin{figure}
\includegraphics[width=9cm]{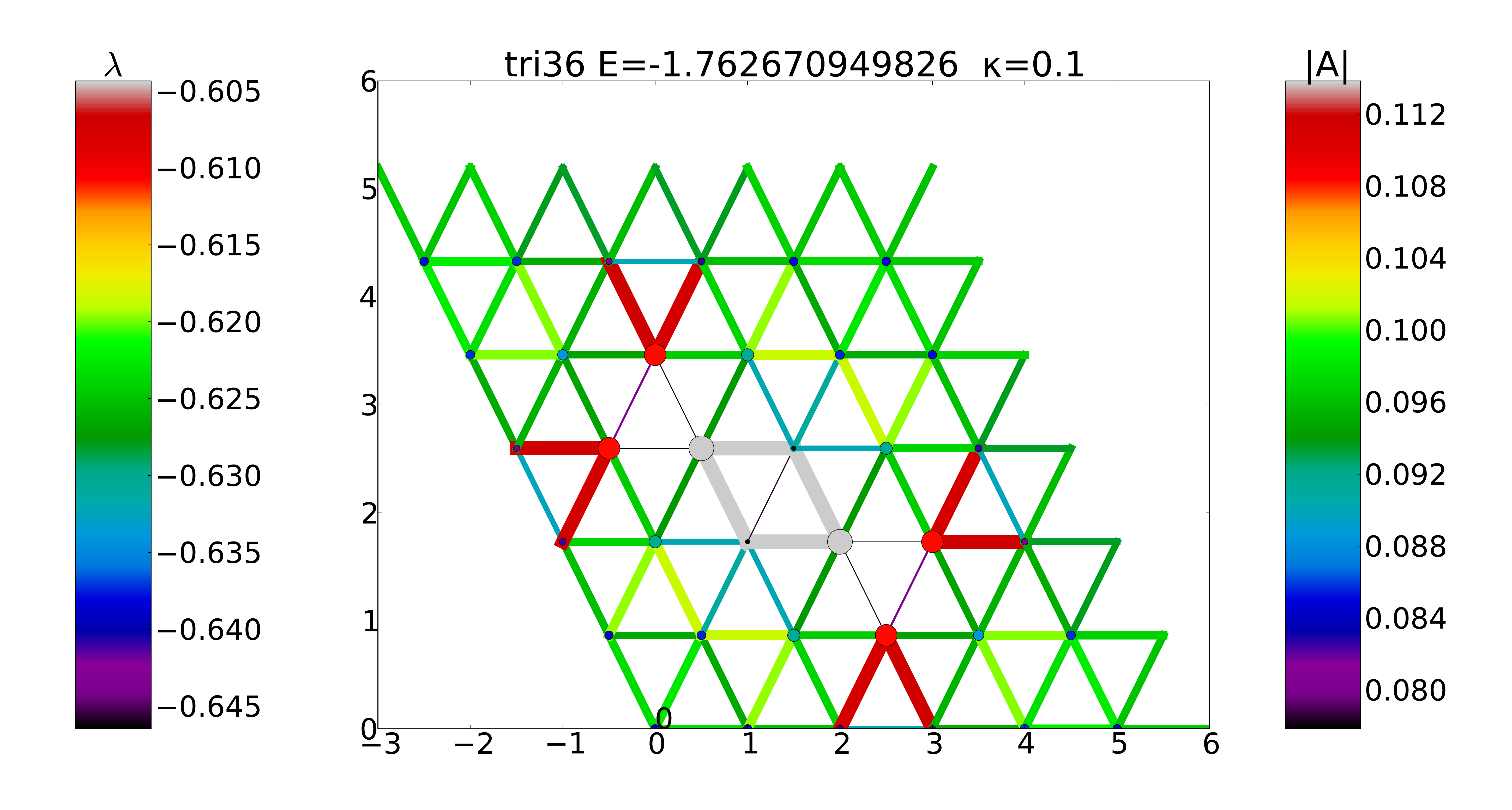}
\includegraphics[width=9cm]{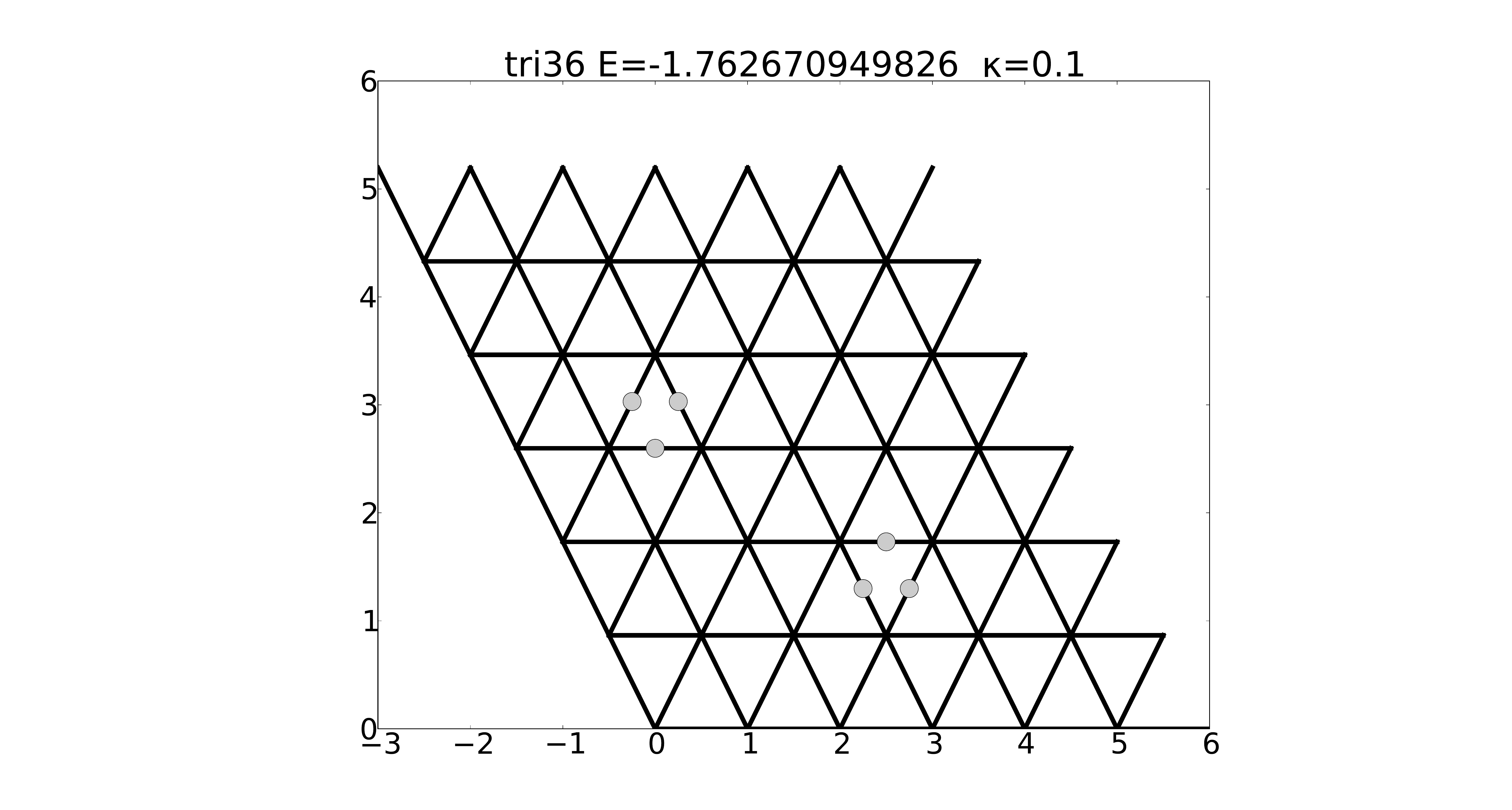}
\caption{(Color online) Top Modulus $|A|$ and chemical
potential for a saddle point with two visons at distance $d=5\sqrt{3}/3$, ($6\times6$ triangular lattice, $\kappa=0.2$). Bottom: the flux vanishes on all diamonds except those with the diagonal tagged with a grey circle. The visons are localized on the triangles with three such circles.}
\label{fig:visons_tri36}
\end{figure}

\begin{figure}
\includegraphics[width=9cm]{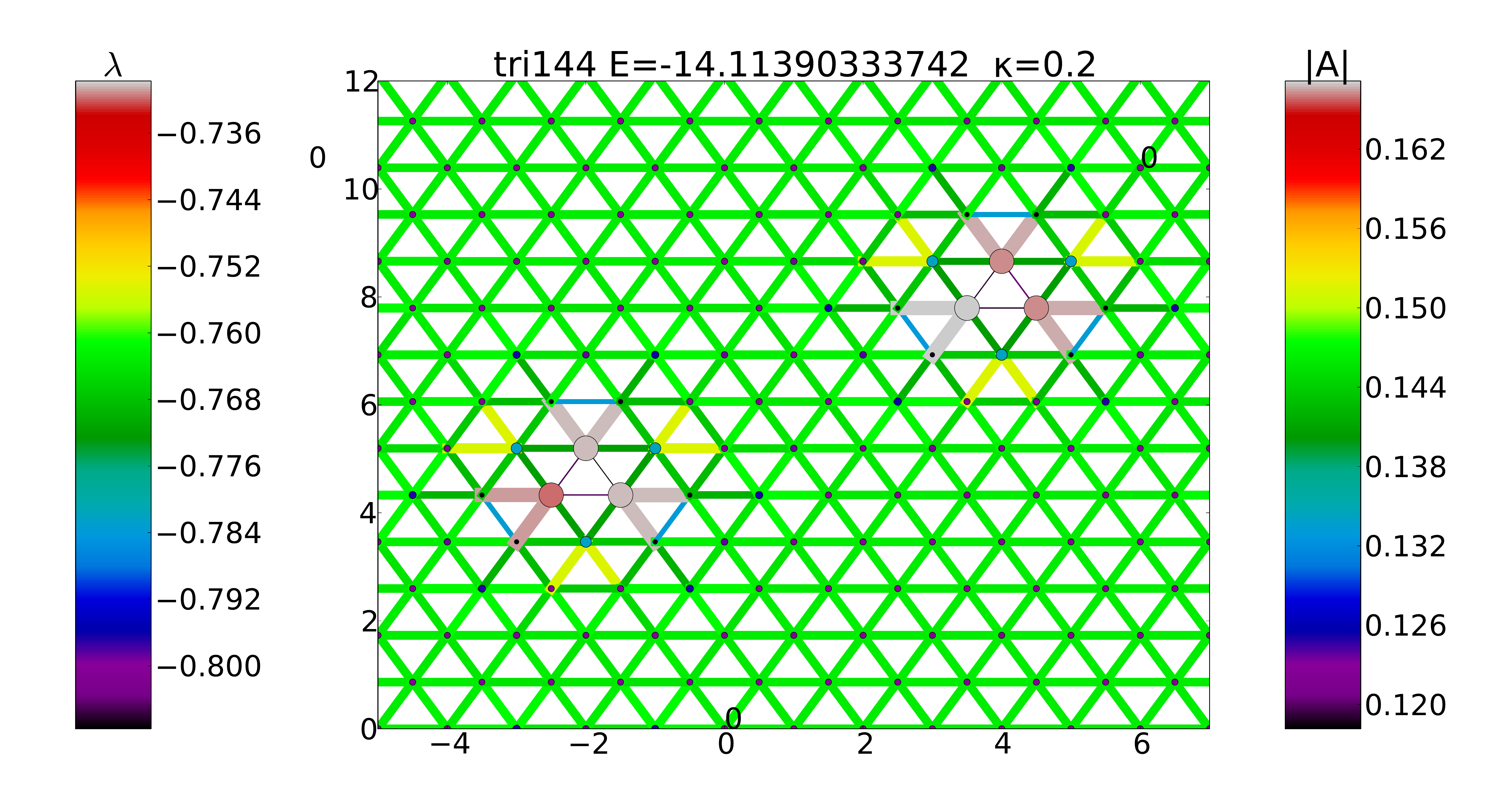}
\caption{(Color online) Modulus $|A|$ and chemical
potential for a saddle point with two visons at distance $d=4\sqrt{3}$, ($12\times12$ triangular lattice, $\kappa=0.2$). A string of reversed $A_{ij}$ (not shown) goes from
the first vison core to the second.}
\label{fig:visons_tri144}
\end{figure}

Due to the presence of flux $\pi$ for each loop encircling a vison core, such a trial state is in general not self-consistent ($\langle\hat A_{ij}\rangle\ne A_{ij}$ and $\langle\hat n_i\rangle\ne \kappa$).
To obtain a self-consistent state (saddle point) the bond moduli as well as the chemical potentials should be re-adjusted in the vicinity of the vison cores.
We obtain numerically such self-consistent vison-pair states by looking at the nearest saddle point in the vicinity of the trial state (second stage of the algorithm described
in Sec.~\ref{ssec:obp}).
These gives access to some energetics of the visons, and to their energy as a function  distance in particular.

Typical vison-pair solutions are displayed in Fig.~\ref{fig:visons_tri36} and Fig.~\ref{fig:visons_tri144}. The modulation  of $|A|$ in the vicinity of each vison is clearly visible, as well as the
higher (less negative) chemical potentials in the core regions. In this case ($\kappa=0.2$) the vison core radius is of the order of two lattice spacings.
These excited states are however not local minima, but saddle points. In all the cases considered here (vison distance from 2 to $4\sqrt{3}$) we find four negative Hessian eigenvalues, with a rather weak dependence on the vison separation.

A calculation of the bond modulations in a vison mean-field state was recently carried out by Huh, Punk and Sachdev\cite{hps11} using and effective model valid in the limit of large spinon gap. Their
result indicates that $|A_{ij}|$ decreases on all the bonds close to the vison core. In our calculation it appears
that the some moduli are indeed depressed,
but some are also enhanced (red bonds in Fig.\ref{fig:visons_tri36}
and grey bonds in Fig.~\ref{fig:visons_tri144}).

\begin{figure}
\includegraphics[width=5cm,angle=-90]{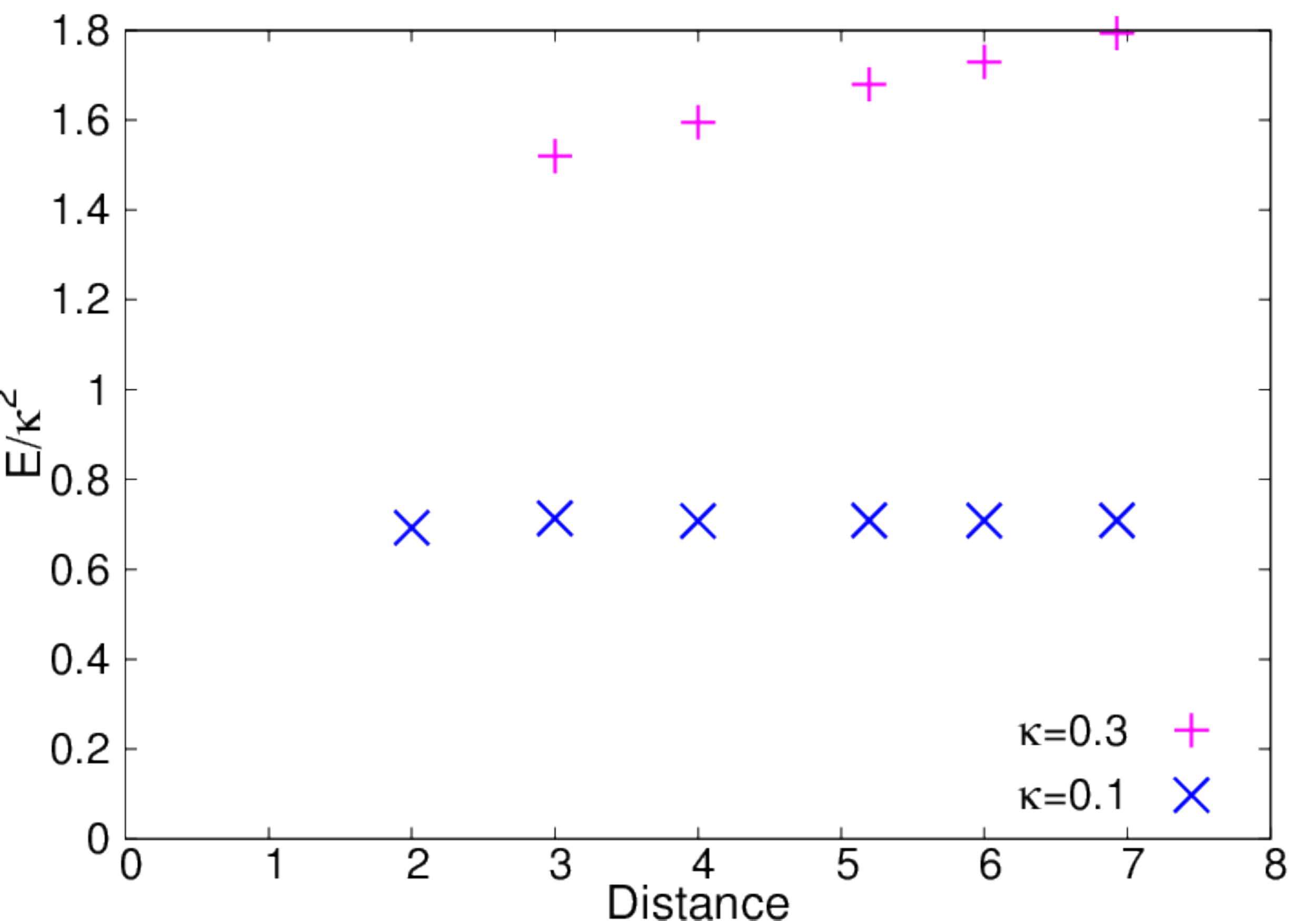}
\includegraphics[width=5.1cm,angle=-90]{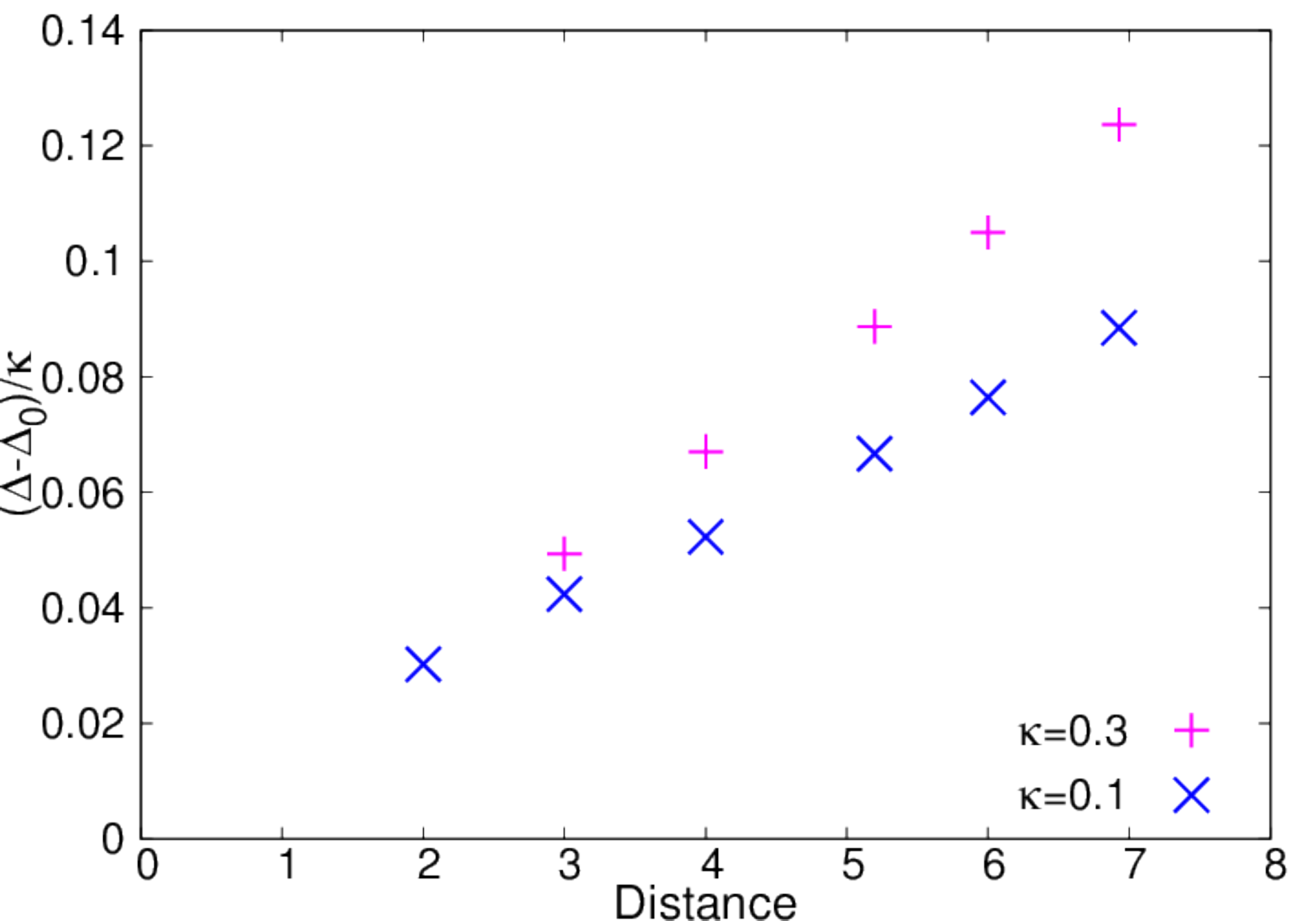}
\caption{(Color online) top: Energy cost of a pair of visons as a function of distance in a 144-site triangular cluster.
Bottom: difference between the spinon gap $\Delta$ in presence of the two visons, and the gap $\Delta_0$ in the absence of visons.
This difference has been scaled by $\kappa$ to compare $\kappa=0.1$ and $\kappa=0.3$). Since $\Delta>\Delta_0$, there is
some {\it repulsion} between a spinon and a vison pair.}
\label{fig:visons_tri144b}
\end{figure}

Some aspects of the visons energetics are summarized in Fig.~\ref{fig:visons_tri144b}. In the large-$N$ framework, these energies should be multiplied by a factor $N$.
It can be checked that their mutual interaction is very weak, since the total energy hardly depends on the
distance, as expected in a gapped $\mathbb{Z}_2$ liquid phase. These calculations also provide some information about the spinon-vison interactions.
This question is important since vison-spinon bound states have fermionic mutual statistics.\cite{rc89}
It appears however that the spinon gap $\Delta$ is slightly {\it higher} in presence of a vison pair than in the ground state ($\Delta_0$).
This indicates some vison-spinon repulsion and makes unlikely the existence
of a bound-state between these two excitations.

\begin{figure}
\includegraphics[width=9cm]{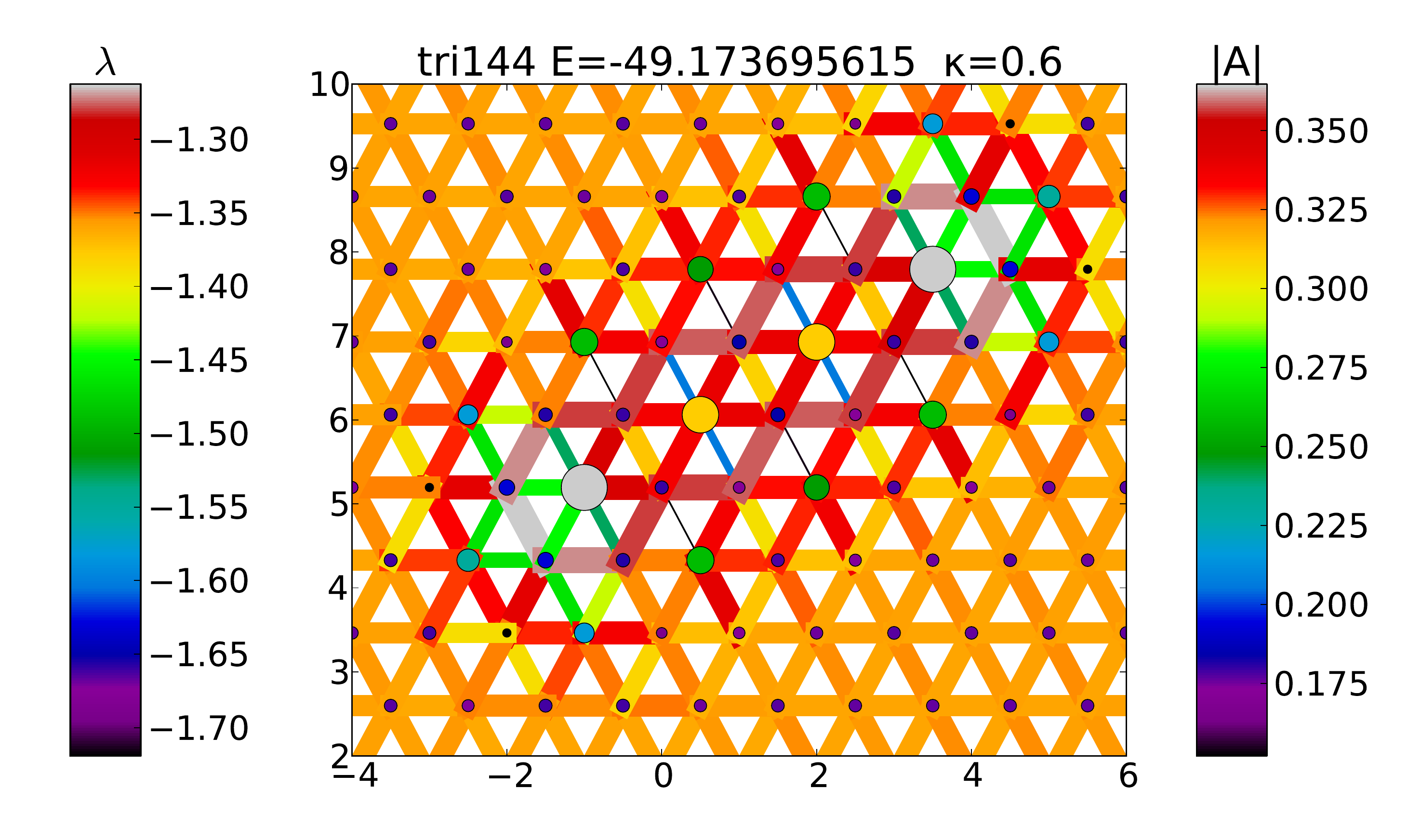}\\
\hspace*{-1cm}\includegraphics[width=8cm]{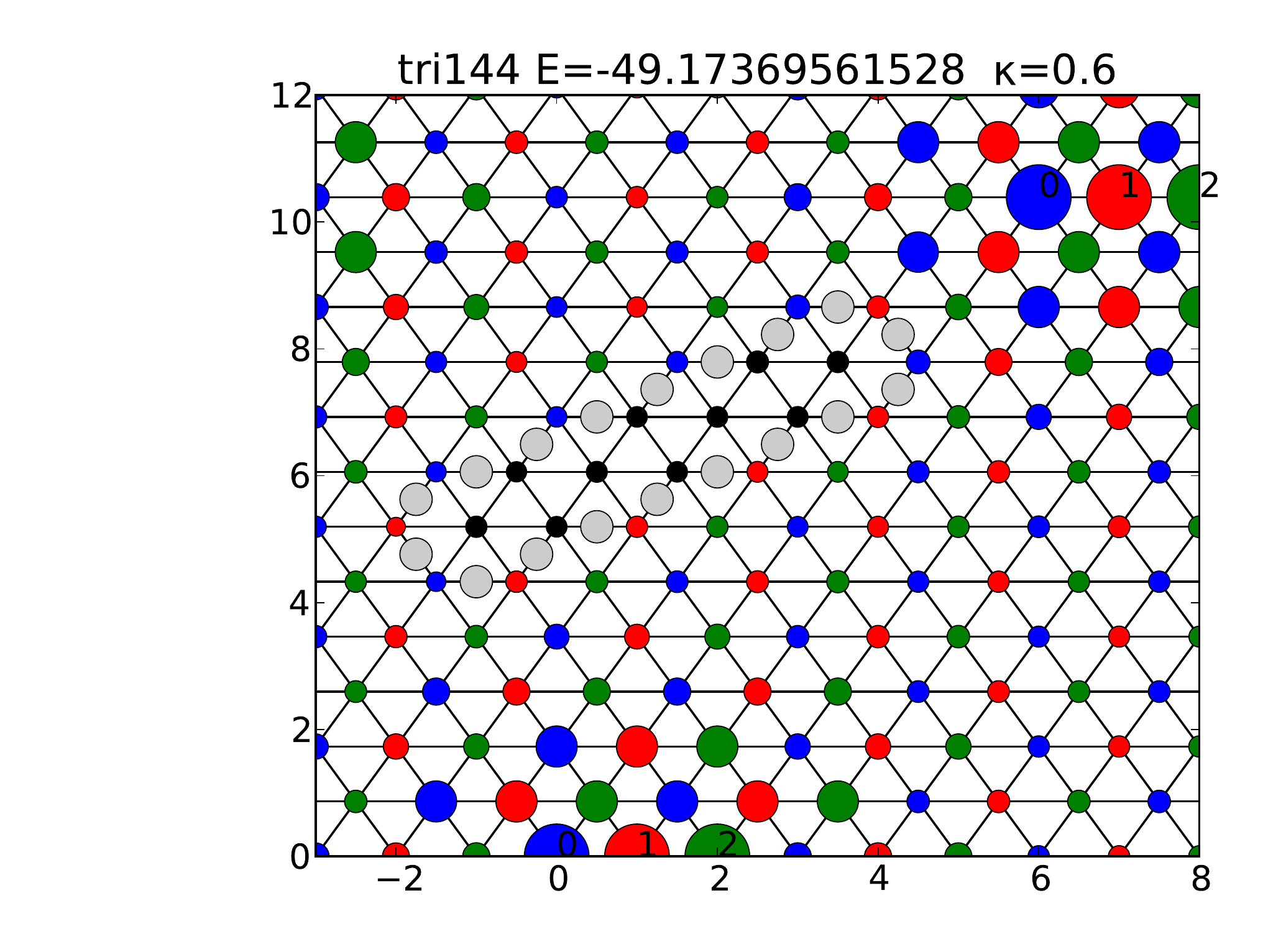}
\caption{(Color online) Highly excited solution on the triangular lattice (144 sites) with $\kappa=0.6$ (ordered phase). Top: bond amplitudes. Bottom:
Spin-spin correlations and fluxes. 
If $\langle S_0\cdot S_i\rangle$ is positive while $\langle S_1\cdot S_i\rangle$ and $\langle S_2\cdot 
S_i\rangle$ are negative, the site $i$ is said to belong to the ``0'' sublattice (blue circles).
Likewise, the site which are positively correlated with site $1$ (resp. $2$) are marked in red (resp. green).
The other sites are  marked in black and the radius of the circle is proportional to the maximum of $|\langle S_0\cdot S_i\rangle|$,
$|\langle S_1\cdot S_i\rangle|$ and $|\langle S_2\cdot S_i\rangle|$.
The diamonds have vanishing flux except for those marked by a grey circle, which have a flux $\pi$ (see text).}
\label{fig:tri144_vortices}
\end{figure}

We investigated the effect of inserting two $\pi$ fluxes far apart in the magnetic phase of the model.
Since the system is magnetically ordered, visons are no longer low-energy excitations.
Starting from a  vison-pair trial state we look for the nearest self-consistent mean field state.
A typical result is shown in Fig.~\ref{fig:tri144_vortices}, where the initial state was chosen to have two localized visons
at the same locations as in Fig.~\ref{fig:visons_tri144}. At the end of the numerical optimization (stage 2 only), it appears the algorithm has converged to
a (unstable) saddle points where some additional pairs diamonds with flux $\pi$ are present.
These additional fluxes form an elongated ring enclosing the two initial vison cores.
Contrary to the vison-pair states in the liquid phase, the spin-spin correlations are strongly modified
all the way inside the ring (bottom panel of \ref{fig:tri144_vortices}).
Indeed, the three sublattice is destroyed, although the spin-spin correlations remain large (black circle radii). This
state appears to be similar to a classical vortex/anti-vortex pair.

\section{Summary and conclusions}

We have developed a numerical method to explore the low-energy SBMFT solutions on finite clusters up to one hundred bonds, without any assumption
on the symmetries of the solutions. The algorithm is able to determine the global energy minimum, the spectrum of its Hessian matrix,  as well as excited saddle points.
The high numerical accuracy allows to resolve saddle points with small energy differences and wery weak spatial modulations.
The Bogoliubov spectrum of {\it spinons} excitations has already been discussed at length in the litterature
on SBMFT. In this work we instead focused on the {\it non-magnetic} excitations associated to
small (quadratic) bond fluctuations in the vicinity of the ground state, or those which correspond to excited saddle points.

At low $\kappa$ the SBMFT describes spin liquids with gapped spinons. In the square lattice case
our calculations confirmed that some low-energy non-magnetic excitations are associated to gauge degrees of freedom.
These excitations  are gapless, and linearly dispersing U(1) ``photons''. They are associated to the first eigenvalues of the Hessian describing small
amplitude phase fluctuations in the vicinity of the ground state.
As expected we observe that these photons get gapped when entering the magnetically ordered phase, due to spinons condensation.
On the triangular lattice, we found saddle points corresponding to
pairs of $\mathbb Z_2$ vortices (visons). We presented some results concerning the energetics of these visons (gap and and weak mutual attraction).

In addition to these excitations wich are qualitatively well undertsood, the SBMFT energy landscape revealed in all cases a large number of low-energy excited saddle points
which do not appear to correspond to some intuitively simple excitation. For instance, on the triangular lattice, the presence of low-energy saddle points with complex/chiral fluxes
do not have a simple explanation in terms of the $\mathbb Z_2$ gauge degrees of freedom which are expected to describe the low-energy physics of a short-range RVB spin liquid.
This surprising observation clearly deserves further investigations.
Can they be related to some real {\it spin} excitations or are they specific to the $N=\infty$ limit ? Can they provide some information about the finite $N$ fluctuations ?

In the magnetic phases of the square and triangular lattice models, the first excited saddle point turn out to be a point-like object, with some
spin ``texture'' localized around some core. This texture appear to be planar in the triangular case and non-planar in the square lattice case, but here also
the precise connnection with spin excitations of an $SU(2)$ model is not obvious. The present study is probably just a first descriptive step toward a better understanding
of this large $N$ limit.

We finally mention that, on the kagome lattice, this approach also reveals a complex landscape with tiny energy scales, and some unexpected
symmetry breaking in the (mean-field) ground state of small clusters. Work is in progress to determine the actual ground state
symmetry on larger clusters (108 sites in particular), where none of the two well-studied states ($\sqrt 3\times \sqrt 3$ and $q=0$) is the the lowest-energy state
for $kappa=1$.\cite{gmSBMFT}

{\it Acknowledgments --- }
I wish to thank Laura Messio, Claire Lhuillier, Vincent Pasquier, Roderich Moessner and Shivaji Sondhi for useful discussions. Part of the calculations
were performed on the computers {\tt titane} and {\tt airain} at the {\it Centre de Calcul Recherche et Technologie} of the CEA.


\begin{thebibliography}{99}
\bibitem{noteSpn}
This mean field Hamiltonian is the same as that obtained
in the $N\to\infty$ limit of the Sp($N$) generalization of the Heisenberg model.\cite{rs91}
\bibitem{note2}
From Eq.~\ref{eq:HA} we have
$\partial E /\partial \lambda=\kappa-\langle \hat n_i \rangle$
and $\partial^2 E /\partial \lambda^2=-\partial \langle \hat n_i \rangle / \partial \lambda$.
Using first order perturbation theory the last expression can be shown to be negative.
\end{thebibliography}
\end{document}